\documentclass[12pt]{article}
\usepackage{graphicx} 
\usepackage{float}
\usepackage{hyperref}
\usepackage{comment}
\usepackage{amssymb}
\usepackage{amsmath} 
\usepackage{bm}
\usepackage{color}

\newcommand{\nsims}{100}

\usepackage{caption}
\usepackage{subcaption}
\usepackage{mathabx}
\usepackage{authblk}
\usepackage{todonotes}
\usepackage{url}
\usepackage{ulem}

\usepackage{subcaption}
\newcommand{\pkg}[1]{{\fontseries{b}\selectfont #1}} 
\newcommand{\At}[1]{A_{#1}}
\newcommand{\A}{\tilde{A}}
\newcommand{\w}{\tilde{\omega}}
\newcommand{\wt}[1]{\omega_{#1}}

\newcommand{\mt}[1]{\mu_{#1}}

\usepackage{nameref}

\usepackage{cite}

\begin{document}


\title{Role Detection in Bicycle-Sharing Networks Using Multilayer Stochastic Block Models}

\author[1]{Jane Carlen$^{}$\footnote{Jane Carlen, Jaume de Dios Pont, and Cassidy Mentus are joint lead authors of the paper.}}
\author[2]{Jaume de Dios Pont$^{*}$}
\author[2]{Cassidy Mentus$^{*}$}
\author[2]{Shyr-Shea Chang}
\author[3]{Stephanie Wang}
\author[4]{Mason A. Porter\footnote{E-mail address of corresponding author: \url{mason@math.ucla.edu}}}
\affil{Data Science Initiative, University of California, Davis\footnote{Now at the \textit{Los Angeles Times}}}
\affil[2]{Department of Mathematics, University of California, Los Angeles}
\affil[3]{Department of Computer Science and Engineering, University of California, San Diego}
\affil[4]{Department of Mathematics, University of California, Los Angeles and Santa Fe Institute}

\maketitle

\begin{abstract}
In urban spatial networks, there is an interdependency between neighborhood roles and the transportation methods between neighborhoods. In this paper, we classify docking stations in bicycle-sharing networks to gain insight into the human mobility patterns of three major United States cities. We propose novel time-dependent stochastic block models (SBMs), with degree-heterogeneous blocks and either mixed or discrete block membership, which classify nodes based on their time-dependent activity patterns. We apply these models to (1) detect the roles of bicycle-sharing docking stations and (2) describe the traffic within and between blocks of stations over the course of a day. Our models successfully uncover work, home, and other districts; they also reveal activity patterns in these districts that are particular to each city. Our work has direct application to the design and maintenance of bicycle-sharing systems, and it can be applied more broadly to community detection in temporal and multilayer networks {with heterogeneous degrees}.
\end{abstract}



\section{Introduction}


Transportation systems and commuting patterns shape and reveal the functional regions in a city \cite{barbosa2018human,barthelemy2019}. The continuing development of network-analysis methods and the increasing availability of transportation data (from a multitude of transportation modes) gives exciting opportunities to improve understanding of urban dynamics at both large and small scales. One prominent example is bicycle-sharing systems, which are an emerging vehicle of urban transportation that can adapt quickly to the needs of travelers. The number of bicycle-sharing programs worldwide has grown rapidly, from $5$ in 2005 to $1571$ in 2018 \cite{schmidt2018ehp}. Over 50 systems were launched in the United States alone from 2010 to 2016, and over 20 bicycle-sharing systems have been launched in France since 2005 \cite{nacto2017bikeshare, etienne2014velib}. Many existing bicycle-sharing systems are also growing. For example, the number of stations in New York City's `Citi Bike' system has more than doubled since it began in 2013. Docked bicycle-sharing systems follow a general structure: Groups of bicycles are parked at `stations' (which are also called `docks' or `hubs') throughout a coverage area, and users withdraw and return bicycles to these stations on demand, with a cost that depends on usage time. A growing portion of bicycle-share systems are dockless (as are the increasingly prominent e-scooters), so users can park bicycles at any location in a coverage area. In the present paper, we analyze docked systems, but we consider how to adapt our models to dockless systems in Section \ref{Discussion}.

Data from bicycle-sharing {systems are available from many cities throughout the world \cite{austwick2013structure,munozmendez2018community,romanillos2016madrid,romanillos2018pulse,wergin18gps}. Such data, which include detailed temporal records and GPS-tracked routes in some cases, are helpful for capturing commuting behavior \cite{fishman2013bike}. Therefore, it is valuable to analyze bicycle-sharing data {to} increase {our} understanding of urban flows and the properties of human commuting. Bicycle-sharing is used often for `last-mile' transportation, bridging the gap between public transportation and a final destination \cite{griffin2016lastmile}{,} and insights into the dynamics of bicycle-sharing systems can help transit systems evolve to meet the needs of changing cities \cite{ashqar2019counts,natera2019multiplexbike}. Studying bicycle-sharing systems is also helpful for maintaining and expanding them \cite{ashqar2020network}. For example, it can facilitate efficient redistribution of bicycles, a question that has received much research attention \cite{shu2013redistribution, prfrommer2014redistribution, singhvi2015redistribution, forma2015redistribution}.
 
In the present paper, we propose two models of temporal network connectivity to capture the functional roles of bicycle-sharing stations. We use the lens of mesoscale-structure detection in time-dependent networks \cite{holme2012temporal,holme2015,fortunato2016community}. We examine trip histories from bicycle-sharing systems in the form of multilayer networks \cite{kivela2014,nutshell2019,Porter2018} in which each layer is a network of trips {that start} in a given hour. 

We aim to partition each of our networks based on a relational equivalence of nodes (a perspective with a rich history in the social-networks literature \cite{lorrain1971struct,rossi2015}), rather than on high internal traffic within sets of nodes \cite{munozmendez2018community}. Data that has been aggregated over long periods of time can shed light on `community' structure in the latter sense \cite{austwick2013structure} through a partition of a network into contiguous spatial clusters \cite{munozmendez2018community}. However, it ignores how bicycle-sharing usage relates to travel patterns during a day \cite{fishman2013bike}. By contrast, our models are designed to detect functional roles of bicycle-sharing stations based on time-dependent behavior. 

The models that we introduce in this paper are time-dependent extensions of stochastic block models (SBMs) \cite{snijders1997sbm, nowicki2001sbm,karrer2011stochastic,peixoto2017bayesian}. We include parameters to describe intra-block and inter-block traffic for each hour, but we fix the block assignment of each node over time. That is, we treat a bicycle-sharing network as a temporal multilayer network with fixed node identities across layers \mbox{\cite{kivela2014,nutshell2019}}. We use the terminology `time-dependent' to emphasize the time-dependent nature of the application of interest and associated interpretations of our results, but our models are relevant for any multilayer network with {the same set of nodes across layers} and no interlayer edges. We also note that the layers do not need to be ordered.

We introduce mixed-membership and discrete-membership versions of our model, where nodes can be members of multiple blocks or exactly one block, respectively. Both versions of our model are degree-corrected. In an SBM without degree correction, the expected weight of an edge is determined by the block {memberships of its incident nodes.} By contrast, in a degree-corrected SBM, the expected {weight of an edge also depends} on the activity levels (i.e., the degrees) of its incident nodes. We extend the degree correction introduced by Karrer and Newman in \cite{karrer2011stochastic} to our multilayer SBMs (see Section \ref{Model}). Incorporating degree correction when modeling networks with heterogeneous degrees can increase the performance of a model while only slightly increasing its complexity \cite{karrer2011stochastic}. 

In Section \ref{Results}, we illustrate that degree correction enables us to identify blocks based on their functional roles in a network, rather than based on their levels of activity. This is especially important for bicycle-sharing networks in which stations have heterogeneous numbers of parking spaces for bicycles and different neighborhoods have different baseline levels of bicycle usage. Our models are applicable to both directed and undirected networks, but we consider only directed examples in the present paper. Additionally, although our models are inspired by the analysis of bicycle-sharing systems, they are applicable more generally to multilayer networks in which nodes have heterogeneous degrees and belong to fixed classes.

Community detection in multilayer networks is a very active research area, and it has yielded insights into many problems in diverse disciplines \cite{bazzi2016generative,valdano2015analytical,cranmer2015kantian,barbillon2017stochastic, kobayashi2019nature,papadopoulos2018network}. Many algorithms for community detection have been generalized to multilayer networks. See, for example, \cite{mucha10multiplex,paul2016null,valles2016multilayer,stanley2016clustering,barbillon2017stochastic,yang2011dsbm, zhang2017random,jeub2017}.

The SBMs that we introduce have parameters that describe block-to-block activity that are distinct for a given period of time; however, block-membership (i.e., community-membership) parameters are fixed across all time periods. Most related SBMs that have been introduced differ in one or both of these conditions. Yang et al. \cite{yang2011dsbm} introduced a discrete-membership SBM in which the parameters for block-to-block activity are fixed over time but block memberships evolve over time. Xu and Hero \cite{xu2014dsbm} and Matias et al. \cite{matias2017dynsbm} proposed related models (for unweighted and weighted networks, respectively). However, unlike the model in Yang et al., their models do not fix block-to-block activity parameters over time. Zhang et al. developed a time-dependent block model with degree correction in which block membership does not change over time and the edge dynamics are described by a continuous-time Markov process, such that edges are added or removed between block pairs at constant rates over time \cite{zhang2017random}. Mixed-membership SBMs with time-evolving communities have also been developed \cite{xing2010dmsbm, ho2011dmsbm}. See Rossetti and Cazabet for a survey of community-detection methods for time-dependent networks \cite{rossetti2018dynamicsurvey}.

Community detection has been applied to urban bicycle sharing using a variety of approaches \cite{borgnat2011shared, austwick2013structure, rosvall2009infomap, munozmendez2018community, xie2018examining, akbarzadeh2018bike, kobayashi2019nature, he2019divvytemporal}. Zhu et al. \cite{zhu18multi} applied $k$-means clustering to undirected, time-dependent usage data from bicycle-sharing systems and other urban systems in New York City. Austwick et al. \cite{austwick2013structure} examined modularity optimization with a directed and spatially-corrected null model to identify communities of stations in several cities. However, they detected communities in time-aggregated data, and their discussion pointed out that there are significant limitations when examining community structure while ignoring time-dependent behavior for bicycle-sharing applications. Munoz-Mendez et al.\ \cite{munozmendez2018community} identified communities by hour for bicycle-sharing data from London using an {\sc InfoMap} algorithm \cite{rosvall2009infomap} that respects the directed nature of edges in the underlying trip networks. The changes that they discovered in communities over time highlight the importance of time of day in the usage of bicycle-sharing systems.

There is some recent work that is closely related to ours. Matias et al. \cite{matias2017semi} constructed a time-dependent, discrete-membership SBM with fixed blocks over time and applied it to bicycle-sharing networks in London. They detected some functional blocks, but their approach does not incorporate degree correction. Xie and Wang \cite{xie2018examining} employed an approach that does not use an SBM directly, yet they were able to successfully partition a bicycle-sharing network to find home and work roles of bicycle-sharing stations. They used the ratio of in-degree to out-degree at different times to discover home--work splits during peak commute times. In the present paper, we observe such splits using {our two-block models}. A similarity of their approach to ours is that it corrects for degree; a key difference is that they relied on human supervision to determine peak hours, whereas our models implicitly increase the weights of more-active time periods in our likelihood function. Etienne and Latifa \cite{etienne2014velib} used a Poisson mixture model to cluster bicycle-sharing stations in Paris based on their time-dependent usage profiles. They were able to capture time-dependent activity for each group, distinguish between incoming and outgoing activity, control for the overall activity level of a given station (via degree correction), and associate identified groups with their role in the city. A major difference between their approach and ours is that we distinguish activity between blocks, which allows us to detect behavior like last-mile commuting that occurs within blocks. We also consider mixed membership, which they proposed as a subject for future work. 

Our paper proceeds as follows. In Section \ref{Data}, we list our data sources and present basic statistical analysis of the data. In Section \ref{Model}, we introduce the two versions of our time-dependent SBM --- a discrete model and a mixed-membership model --- and we show that they are equivalent up to a constraint. In Section \ref{Computation}, we describe the estimation algorithms for our discrete and mixed-membership SBMs. In Section \ref{SimStudy}, we generate synthetic networks using both of our models and a related model by Matias et al.~\cite{matias2017dynsbm}, and we evaluate {the} fits of our models and theirs to such generated data. In Section \ref{Results}, we present the results of our models for the bicycle-sharing networks in Los Angeles, San Francisco, and New York City. We discuss the implications of our work for bicycle-sharing systems and suggest areas of further study in Section \ref{Discussion}. We show some additional details of our work in the \nameref{appendix}. We provide code and data to implement our models and replicate the results in our paper at \url{https://github.com/jcarlen/tdsbm_supplementary_material}.


\section{Data}\label{Data}

We examine United States bicycle-sharing systems in Los Angeles, the Bay Area, and New York City. For Los Angeles, we study only the system's downtown region, which is self-contained in the sense that very few trips connect it to stations that are not downtown. Similarly, for the Bay Area, we consider only the San Francisco network and exclude stations in Mountain View, Palo Alto, and San Jose. (If we applied our models to these nearly disconnected components, it would be better to treat them as separate networks than to model them jointly.) Each of our three focal systems covers a dense urban area, but they vary in their sizes and daily usage. Because of this variation and how the data were reported, we study different time periods for each system. We also selected our time periods to exclude days that are likely to be extremely hot or extremely cold. Each of the bicycle-sharing systems that we study has an open-data portal, from which we downloaded our data. After cleaning (see our discussion in the next paragraph), our data consist of the following: 
\begin{itemize}
\item Downtown Los Angeles: October--December 2016; there are 61 stations and 40,130 trips, of which 73.4\% are during weekdays \cite{citibikeData}.
\item San Francisco: September 2015--August 2016; there are 35 stations and 267,412 trips, of which 92.1\% are during weekdays \cite{bayareaData}.
\item New York City: October 2016; there are 601 stations and 1,551,692 trips, of which 75.6\% are during weekdays \cite{metrobikeData}.
\end{itemize}

A trip consists of a user checking out a bicycle from a fixed location (a station that includes multiple parking spaces) and returning it to a station. The data for each trip include the starting time; the ending time; and the starting and ending locations by station ID, latitude, and longitude. Each data set also has a few additional fields about the users; these details include whether they have memberships in the bicycle-sharing system, but we do not use this information in our investigation. We cleaned the data by removing anomalous trips, including extremely short and extremely long trips,\footnote{We take extremely short trips to be those that last two minutes or less. We take extremely long trips to be those that last 90 minutes or more in Los Angeles and San Francisco and 120 minutes or more in New York City (because of the larger coverage areas of stations in New York City,).} and trips to or from a station that were used for testing or maintenance (as indicated in the data). We also excluded a very small number of stations (two in Los Angeles and six in New York City) that did not have at least one departure and at least one arrival during the given time period. Finally, we excluded one station in New York City that was accessible to other stations only by ferry; it was involved in only nine trips during the given time period. In total, cleaning removed 7.1 \% of the trips in Los Angeles, 4.5 \% of the trips in San Francisco, and 1.4 \% of the trips in New York City. 

We retain self-edges, which represent trips that start and end at the same station. Although it is common to remove self-edges when analyzing networks \cite{newman10networks}, we expect the self-edges in our data to have a very similar data-generating mechanism as trips to nearby stations. As in \cite{karrer2011stochastic}, including self-edges also simplifies some aspects of parameter estimation. When fitting our models, we include only weekday trips, as we observe that weekday and weekend activity follow distinct patterns (see Figure \ref{fig:byhour}). Moreover, weekend activity does not reflect commuting behavior. From the data sets, we construct multilayer networks that are both weighted and directed. In our networks, nodes represent stations, edge values encode the number of directed trips from one station to another that begin in a specified time period, and each of the $24$ layers consists of the trips that start in a certain hour. 


\subsection{Preliminary Data Analysis}

When examining the bicycle-sharing data sets, we discovered strong daily usage patterns, which motivate the development of our time-dependent SBMs. In Figure \ref{fig:byhour}, we illustrate these usage patterns by plotting the number of trips by starting hour for each city. In each city, weekday activity spikes during morning and evening commuting hours (and Los Angeles also has a mid-day spike), whereas weekend trips peak in early afternoon. Similar patterns were observed previously for bicycle-sharing systems in New York and many other cities \cite{austwick2013structure, etienne2014velib, zhu18multi, xie2018examining, aqil14tp19}. 

For individual stations, the morning and evening peaks for arriving trips (i.e., in-edges) and departing trips (i.e., out-edges) are often unbalanced. One direction has a stronger morning peak, and the opposite direction has a stronger evening peak. In the bottom panel of Figure \ref{fig:byhour}, we show an example of this imbalance for the six busiest stations in San Francisco. (In Appendix \ref{appendix:svd}, we further explore this issue using singular-value decomposition (SVD) of the trip data.) These patterns suggest the primary ways in which stations are used in commuting, and they motivate our time-dependent identification of stations into `home' and `work' types.

\begin{figure}\flushleft
\includegraphics[scale=.13]{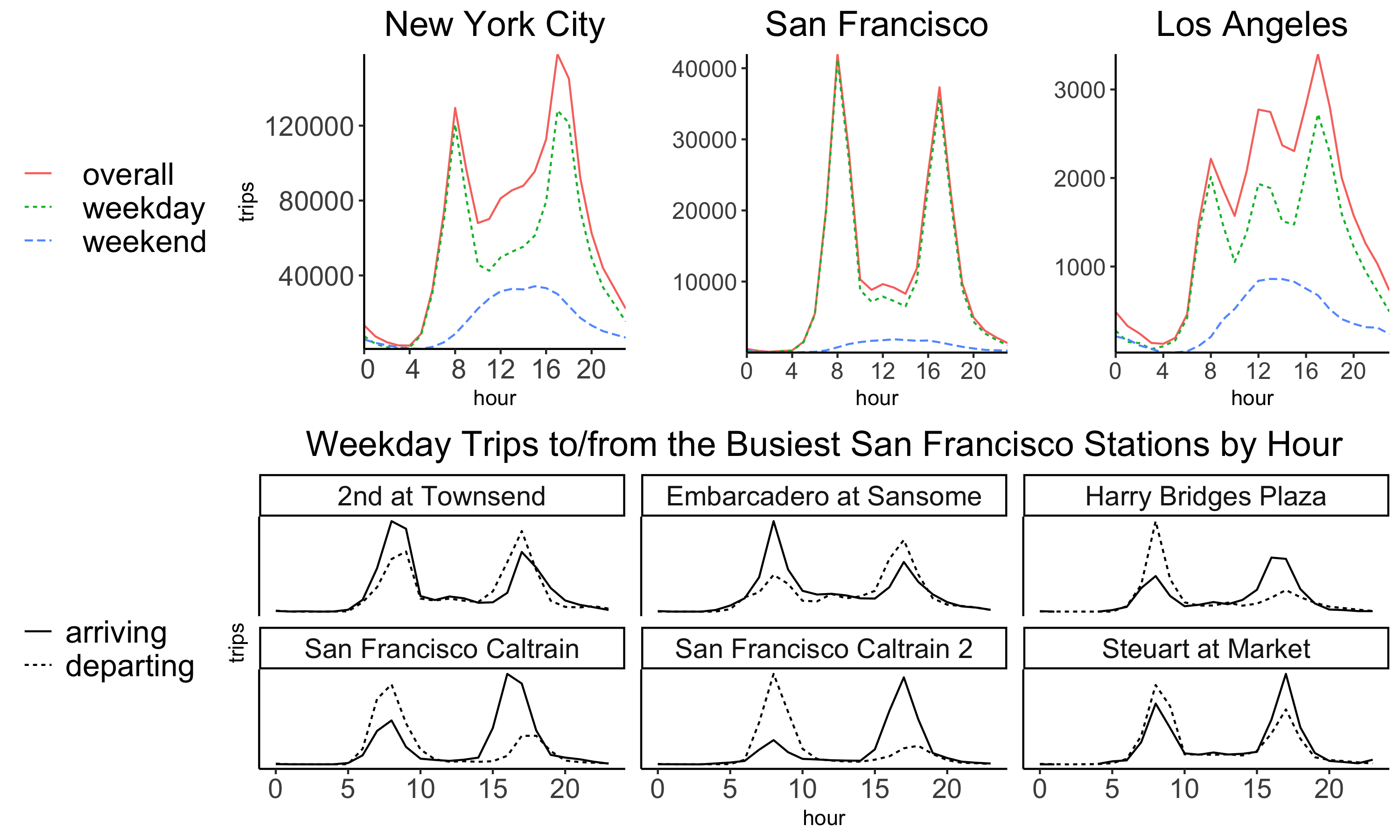}
\caption{(Top) Total bicycle trips by hour for weekdays, weekends, and overall in New York City, San Francisco, and Los Angeles. Hour $0$ designates midnight. (Bottom) Trips that arrive (solid curve) and depart (dashed curve) by hour in the six San Francisco stations with the largest overall trip counts.}
\label{fig:byhour}
\end{figure}

Another characteristic of our data that we incorporate into the design of our models is the strong positive Pearson correlation coefficient between the total (summed over all time periods) in-degree and out-degree of each station: $0.99$ in New York, $0.98$ in San Francisco, and $0.91$ in Los Angeles. Accordingly, we apply degree correction to the overall degree of each station, rather than to the in-degrees and out-degrees separately (see Section \ref{TDMM-SBM}). This correlation is an intrinsic feature of docked bicycle-sharing systems, because at some point bicycles must be returned to a station before any new trips can originate there. (The use of trucks to redistribute bicycles within a system can loosen this requirement.) More generally, the strong correlation also relates to a famous axiom of human mobility: for each current of travel, there is a countercurrent \cite{barbosa2018human}.



\section{Our Stochastic Block Models}\label{Model}

In this section, we introduce our time-dependent mixed-membership stochastic block model (TDMM-SBM) and time-dependent discrete stochastic block model (TDD-SBM).

Stochastic block models are a popular class of statistical network models \cite{peixoto2017bayesian}. The motivating principle of SBMs is a notion of \textit{stochastic equivalence} in which edges whose incident nodes have the same block membership are identically distributed. It is a standard assumption of SBMs that edge {weights} are independent, given the block memberships of its nodes. More formally, an unweighted (i.e., Bernoulli) random graph $Y$ (with adjacency matrix $A$) from an SBM with $K$ blocks is defined by
\begin{equation}\label{sbm_model}
	P(A_{ij} = 1|G) = \eta_{g_i g_j}\,,
\end{equation}
where $G$ (with components $g_i\in \{1,2,\ldots, K\}$) is a vector of block assignments for the nodes of $Y$ and $\eta$ is a $K \times K$ matrix of block-to-block edge probabilities. This definition allows directed graphs, in which $\eta$ can be asymmetric. For early presentations of SBMs, see \cite{holland1983sbm, faust1992blockmodel, snijders1997sbm, nowicki2001sbm}. More recent advances have added flexibility to SBMs. Examples include mixed-membership and overlapping-membership SBMs \cite{airoldi2008mixed, latouche2011overlap}, models with covariates \cite{blockmodelsR}, degree-corrected SBM{s} \cite{karrer2011stochastic}, and Bayesian implementations \cite{peixoto2017bayesian}. Applications of SBM to {time-dependent} networks include (1) discrete-membership \cite{yang2011dsbm,zhang2017random} and mixed-membership \cite{xing2010dmsbm, ho2011dmsbm} versions in which nodes can change blocks and (2) discrete-membership versions with fixed blocks over time, but without degree-correction \cite{matias2017semi, xu2014dsbm}. 


\subsection{Time-Dependent Mixed-Membership Stochastic Block Model (TDMM-SBM)}\label{TDMM-SBM}

We now describe the framework for our mixed-membership SBM. Let $i,j \in \mathcal{N}$ (with $|\mathcal{N}| = N$) be nodes, which represent bicycle stations; let $g,h \in \mathcal{K}$ (with $|\mathcal{K}| = K$) be blocks. We treat both the number of nodes and the number of blocks as fixed and given. Our data is a three-dimensional array of size $(N,N,T)$, where $T$ is the number of time slices (i.e., time layers). We consider hourly groupings of the trips based on their starting times, and we do not include interlayer edges.\footnote{In the bicycle-sharing networks that motivate our work, interlayer edges (beginning at one time slice and ending at another) would help capture the length of trips, but that is already captured implicitly in the distance between station pairs corresponding to each edge. Including interlayer edges is unlikely to add to the insights we gain from our models, but it would significantly increase their complexity.} The quantity $\At{ijt}$ is the observed number of trips from station $i$ to station $j$ with a starting time that is at least $t$ and less than $t+1$. Let $\tilde{A}_{ij}=\sum_{t=0}^{23}\At{ijt}$ denote the {edge} weights of the associated time-aggregated matrix. Our network is a directed multilayer network, so we count each trip that both starts and ends at a node $i$ during hour $t$ (i.e., self-edges) exactly once in $A_{iit}$.

For each node $i$, there is a latent length-$K$ vector of real numbers $C_{ig}\in [0,1]$. These numbers, which we will estimate, represent the mixed-membership block assignment of each node. The block-assignment parameter $C_{ig}$ indicates the `strength' of node $i$ in block $g$. For each ordered pair $g,h$ of blocks and each time $t\in \{0, 1, \ldots, 23\}$ (where $t = 0$ represents the hour that starts at midnight), there is a parameter {$\wt{ght}$, which we call the `block-connectivity' parameter or `block-to-block' parameter,} that represents the directed activity from block $g$ to block $h$ during hour $t$. Note that $\wt{ght}$ need not be equal to $\wt{hgt}$ if a network is directed; in our application, this captures any asymmetries in the numbers of trips with respect to reversing origins and destinations. We also define the notation $\w_{gh}=\sum_{t=0}^{23} \wt{ght}$ for our time-aggregated matrix. 

For each pair of nodes, $i$ and $j$, we assume that the number of trips that depart from $i$ {at time $t$} and arrive at $j$ is Poisson-distributed with mean $\mt{ijt} = \sum_{g,h} C_{ig}\wt{ght} C_{jh}$. Our use of the Poisson distribution follows \cite{karrer2011stochastic} and \cite{peixoto2017bayesian}, facilitates computation, and is standard for modeling count data (although overdispersion is a concern). 

For identifiability, we apply the constraint $\sum_i C_{ig}=1$ for all $g$. This does not constrain the set of possible models in terms of realizable mean edge activities $\mu_{ijt}$. Consider a model with unconstrained parameters $\omega_{ght}$ and $C_{ig}$. 
{The model with parameters $\omega'_{ght}$ and $C'_{ig}$ such that $C'_{ig}=\frac{C_{ig}}{\sum_{j} C_{jg}}$ and
$\omega'_{ght}=\omega_{ght}\left(\sum_{j} C_{jg}\right)\left(\sum_{j}C_{jh}\right)$ is an equivalent {parameterization}, because the means of the distributions of edge weights are equal to the those of the model with unconstrained parameters. That is, $\mu'_{ijt}=\sum_{g,h}C'_{ig}\omega'_{ght}C'_{jh}=\sum_{g,h}C_{ig}\omega_{ght}C_{jh}=\mu_{ijt}$. 

Given that $\sum_{i} C_{ig}=1$, we can think of $C_{ig}$ as the proportion of the total activity of block $g$ from the activity of node $i$. We interpret $C_{ig}$ relative to $C_{ih}$ as how strongly block $g$ is associated with node $i$ relative to how strongly block $h$ is associated with node $i$. The expected total number of trips at node $i$ is $\sum_g C_{ig}\sum_{h,t} (\omega_{ght}+\omega_{hgt})$. Accordingly, $\sum_g C_{ig}$ is a measure of the activity of node $i$ in which we do not weight each $C_{ig}$ term by the total activity of the corresponding block. We use these quantities when plotting the block assignments that we infer in our data using the TDMM-SBM (see the left panels of Figures \ref{fig:LA_mixed_discrete} and \ref{fig:SF_mixed_discrete}) because they help ensure that we do not overlook blocks with important usage patterns but relatively lower activity. The parameter $C_{ig}$ is analogous to the degree-correction parameter for SBMs that was introduced in \cite{karrer2011stochastic}, but we apply it to mixed block membership. We elaborate on this connection in Subsection \ref{TDD-SBM}, where we introduce a model that specifies that nodes have only one block. 

We now compute the likelihood function that we will optimize to obtain the maximum-likelihood estimate (MLE). We assume conditional independence between {the Poisson-distributed} hourly numbers of trips along each edge, given model parameters, so the likelihood of the data is
\begin{align}\label{llik_TDMM-SBM}
	L(G;\mathbf{\omega},\mathbf{C})=
\prod_{t=0}^{23}\prod_{i,j\in\mathcal{N}} \frac{(\mt{ijt})^{\At{ijt}}}{\At{ijt}!}\exp\left(-\mt{ijt}\right)\,,
\end{align}
where $\mathbf{\omega}$ and $\mathbf{C}$ give the model parameters (i.e., $\mathbf{\omega}=\{\omega _{ght}\}$ and $\mathbf{C}=\{C_{ig}\}$). Note that $\mt{ijt}=\sum_{g,h} C_{ig}\wt{ght} C_{jh}$ is a function of these parameters, the set $\mathcal{N}$ of nodes in the network is fixed and given, and the number $K$ of blocks is also fixed and given. The unnormalized log-likelihood is
\begin{align} \label{logroll}
	\ell(G;\mathbf{\omega},\mathbf{C})=
\sum_{t=0}^{23}\sum_{i,j\in\mathcal{N}} 
\left[\At{ijt}\log\left({\mt{ijt}}\right)
- \mt{ijt}\right]\,,
\end{align}
although we omit the addition of the constant $-\sum_{i,j,t}\log\left(A_{ijt}!\right)$ because it does not affect maximum of the function. (In \eqref{logroll} and throughout this paper, we use `log' to denote the natural logarithm.)


\subsection{Time-Dependent Discrete Stochastic Block Model (TDD-SBM)}
\label{TDD-SBM}

We derive a single-membership SBM from our mixed-membership SBM by making the extra assumption that, for each node $i\in \mathcal{N}$, it is the case that $C_{ig}>0$ for only one block $g \in \mathcal{K}$. (This SBM is thus the `discrete version' of our model.) For our single-membership SBM, we introduce some new notation to aid our description and promote consistency with the notation in other work \cite{karrer2011stochastic,zhu2013oriented}. For a given node $i$, the block $g$ for which $C_{ig}>0$ is the block $g_i$ that includes node $i$. We introduce a scalar parameter $\theta_i=C_{ig_i}$ for each node $i$; it is a multilayer extension of the degree-correction parameter in Karrer and Newman \cite{karrer2011stochastic}. The expectation of the Poisson distribution of the value of an edge from node $i$ to node $j$ at time $t$ is $\theta_i\theta_j\omega_{g_ig_jt}=C_{ig_i}\wt{g_{i},g_{j}t} C_{jg_{j}}{=\mu_{ijt}}$. We retain the sum constraints of our mixed-membership model, so $\sum_{i\in g}\theta_i = 1$ for all $g$.

We compute optimal values for the parameters $\mathbf{\omega}$ and $\mathbf{\theta}=\{\theta_i\}_{i\in\mathcal{N}}$. As in our TDMM-SBM, take $\mathcal{N}$ and $K$ to be fixed and given. Again dropping the constant term $-\!\sum_{i,j,t}\log\left(A_{ijt}!\right)$, the log-likelihood of our single-membership SBM is
\begin{align}\label{llik_TDD-SBM}
	\ell(G;\mathbf{\omega}, \mathbf{\theta}) = \sum_{t}\sum_{g,h}\sum_{i\in g, j\in h} \left[\At{ijt}\log\theta_i+\At{jit}\log \theta_j+ \At{ijt}\log \wt{ght} -\theta_i\theta_j \wt{ght}\right].
\end{align} 
This resembles the degree-corrected SBM in equation (14) of \cite{karrer2011stochastic}, but we have adapted it to directed networks and summed over time layers.

We find explicit formulas for the conditional MLEs of $\theta_i$ and $\wt{ght}$, {given the block memberships $g_i$}. In the following calculations, removal of $t$ from the subscript of a parameter and insertion of a tilde designates a sum over all $t$. Specifically, we define $\A_{ij}=\sum_{t=0}^{23} \At{ijt}$ and $\w_{gh}=\sum_{t=0}^{23} \wt{ght}$. We differentiate $\ell$ with respect to $\wt{ght}$ to obtain
\begin{align*}
	\frac{\partial}{\partial \wt{ght}} \ell=
\frac{\sum_{i \in g, j \in h}\At{ijt}}{\wt{ght}}-1\,,
\end{align*}
where we have used the {blockwise} sum constraints on {$\theta$ to simplify $\sum_{i\in g, j\in h}\theta_i\theta_j$ to $1$}. Therefore, the {conditional} MLE for $w_{ght}$ is
\begin{align}\label{w_ght-MLE}
	\hat{\omega}_{ght}=m_{ght}\,,
\end{align}
where $m_{ght}$ is the sum of the weights of the edges from nodes in block $g$ to nodes in block $h$ during hour $t$. That is, $m_{ght} = \sum_{i\in g,j\in h}\At{ijt}$.

We then differentiate $\ell$ with respect to $\theta_i$ to obtain
\begin{align*}
	\frac{\partial}{\partial\theta_i}\ell=\frac{\sum_j \A_{ij}+\sum_{j} \A_{ji}}{\theta_i}
-\sum_{h}\w_{g_i h}-\sum_{h}\w_{h g_i}\,.
\end{align*}
At $\hat{\omega}_{ght}$, the {conditional} MLE for $\theta_i$ is
\begin{align}\label{theta_i-MLE}
	\hat{\theta_i}=\frac{\sum_j \A_{ij}+\A_{ji}}{\sum_{g}\tilde{m}_{gg_i}+\tilde{m}_{g_ig}} =\frac{k_i}{\kappa_{g_i}}\,,
\end{align}
where $k_i=\sum_j \left(\A_{ij}+\A_{ji}\right)$ is the {degree of node $i$ (i.e., the sum of the in-degree and the out-degree\footnote{When describing weighted networks, we use the term `degree' to refer to the weighted degree. We use the terms `in-degree' and `out-degree' to describe the weighted in-degree and weighted out-degree, respectively.} of $i$)} over all time periods; 
 $\tilde{m}_{gh}=\sum_{t=0}^{23} m_{ght}$; and $\kappa_{g}=\sum_{h} \left(\tilde{m}_{gh}+\tilde{m}_{hg}\right)$ is the sum of the in-degrees and out-degrees of all nodes in block $g$ over all time periods. The term $2\tilde{m}_{gg}$ in the equation for $\kappa_g$ implies that we count each intra-block edge twice: once for emanating from $g$ and once for arriving at $g$. Similarly, $k_i$ includes the term $2\tilde{A_{ii}}$, so we count self-edges twice in this term.
 
The parameter MLE $\hat{C}_{ig_i} =\hat{\theta}_i$ in the TDD-SBM for directed, multilayer networks is analogous to the degree-correction parameter in the degree-corrected SBM for undirected networks with one layer \cite{karrer2011stochastic}. Indeed, the MLE for the degree correction parameter of a node in the latter model is the degree of that node divided by the sum of the degrees of all nodes in its block. Another similarity between a degree-corrected SBM and our TDD-SBM is that at the MLE of the TDD-SBM, the expectation of the (time-aggregated) degree of a node $i$ is equal to the degree of node $i$ from the observed data. That is, $\sum_{t}\sum_{j}\left(\mu_{ijt}+\mu_{jit}\right) = k_i$. (See {Appendix \ref{appendix:node_deg}} for the proof.) For our mixed-membership SBM, we are not aware of such a precise relationship between the data and the expected value of model statistics, although there does appear to be a {strong} positive correlation between the time-aggregated node degrees and the sum of the mixed-membership parameters ($\sum_g C_{ig}$ for all $i$).

We now {substitute the above conditional parameter estimates into the unnormalized log-likelihood of the TDD-SBM to obtain
\begin{align*}
	\sum_{t}\left[\sum_{i,j} \left(\At{ijt}\log \left(\frac{k_i}{\kappa_{g_i}}\right)+\At{jit}\log\left(\frac{k_j}{\kappa_{g_j}}\right)\right)+\sum_{g,h} m_{ght}\log m_{ght} - \sum_{g,h} m_{ght}\right]  \\
	=
\sum_i k_i \log k_i - \sum_i k_i\log {\kappa_{g_i}}+\sum_t\sum_{g,h} m_{ght}\log m_{ght} -\tilde{m}\,,
\end{align*}
where $\tilde{m}$ is the total number of edges in the network. By a similar calculation as one in \cite{karrer2011stochastic}, we obtain
\begin{align*}
	\sum_{i} k_i\log \kappa_{g_i}&=\sum_t\sum_{g}  \sum_{i\in g}  k_{it} \log\kappa_g\\
&=\sum_t \sum_{g}\sum_{i\in g}\left( k_{\text{in},it} \log\kappa_g + k_{\text{out},it} \log\kappa_g\right)\\
&=\sum_t\sum_g\kappa_{\text{out},gt}\log\kappa_g+\sum_t\sum_h\kappa_{\text{in},ht}\log \kappa_h\\
&=\sum_t\sum_g\sum_h m_{ght}\log\kappa_g+\sum_t\sum_g\sum_h m_{ght}\log \kappa_h\\
&=\sum_{t}\sum_{g,h} m_{ght}\log \kappa_g\kappa_h\,,
\end{align*}
where $k_{\text{in},it}$ and $k_{\text{out},it}$ are the in-degrees and out-degrees of node $i$ during hour $t$, the quantity $\kappa_{\text{in},gt}=\sum_{i\in g} k_{\text{in},it}$ is the number of edges that arrive at $g$ in hour $t$, and $\kappa_{\text{out},gt}=\sum_{i\in g} k_{\text{out},it}$ is the number of edges that leave $g$ in hour $t$. Including only the terms that depend on block assignments yields the objective function
\begin{align}\label{tdd-sbm-objective}
	\sum_t\sum_{g,h} m_{ght}\log\left(\frac{m_{ght}}{\kappa_g\kappa_h}\right)\,,
\end{align}
which we maximize to obtain the MLE of the block assignments.

Unlike the directed SBM of Zhu et al.~\cite{zhu2013oriented}, we do not have two {degree-correction parameters for each station. Their directed SBM includes a parameter $\theta_i^{\text{in}}$ to modify in-degree and a parameter $\theta_i^{\text{out}}$ to modify out-degree. The validity of using undirected degree-correction parameters $\{\theta_i\}_{i\in\mathcal{N}}$ to model directed networks depends on the presence of a large correlation between the time-aggregated in-degrees and out-degrees of the nodes. In this situation, including distinct $\theta_i^{\text{in}}$ and $\theta_i^{\text{out}}$ parameters would lead to overfitting. We found Pearson correlations between in-degree and out-degree of over $0.9$ for all of the bicycle-sharing networks that we examined (see Section \ref{Data}). 

Our model is able to capture the directedness of travel at a given time because the $\omega_{ght}$ parameters are not constrained to be symmetric. To increase the applicability of our work to other networks (which may not have such large correlations), we provide estimation code in our implementation materials (see Section \ref{Computation}) for a version of our TDD-SBM with directed degree-correction parameters $\theta_i^{\text{in}}$ and $\theta_i^{\text{out}}$.


\section{Computations}\label{Computation}

In this section, we describe the algorithms that we use for the TDMM-SBM and the TDD-SBM. 


\subsection{Inference using the TDMM-SBM}

Let $\mathbf{\omega}=\{\wt{ght}\}$ be the $K\times K\times T$ array of block-connectivity parameters, and let $\mathbf{C}=\{C_{ig}\}$ be the matrix of node-strength parameters. We estimate the model parameters using a two-step gradient descent.\footnote{Although we are maximizing a function and thus technically performing gradient ascent, we refer to this class of method by its more common monicker of `gradient descent'.} First, we move in the direction of the gradient with respect to $\mathbf{\omega}$ and update the block-connectivity parameters. Second, we move along the direction of the gradient with respect to $\mathbf{C}$ and update the node-strength parameters. 

In our algorithm, we let $\omega^{(n)}$ and $\mathbf{C}^{(n)}$ denote the $n^{\text{th}}$ updates of the block-connectivity and node-strength parameters, respectively. We initialize the algorithm with random values $\mathbf{\omega}^{(0)}$ and $\mathbf{C}^{(0)}$ with components distributed according to $\exp(X)$, where $X$ is a Gaussian random variable with mean $0$ and variance $1$. (That is, we draw random values from a log-normal distribution.) We denote the mean activity along edge $(i,j)$ with initial parameters $\mathbf{\omega}^{(0)}$ and $\mathbf{C}^{(0)}$ by $\mu_{ijt}^{(0)}$. We scale the parameters so that the TDMM-SBM at the starting point of the optimization has the same mean number of trips as the data. Specifically, we multiply the block-connectivity parameters $\omega^{(0)}_{ght}$ by $\left(\sum_{t}\omega^{(0)}_{ght}\right)^{-1}\left(\sum_{i,j,t} \At{ijt}\right)/K^2$ and normalize $\mathbf{C}^{0}$ to satisfy the constraint $\sum_i C_{ig}^{(0)}=1$ for each block $g$. This yields $\sum_{ijt} \mu_{ijt}^{(0)}=\sum_{i,j,t} \sum_{g,h} C_{ig}^{(0)}\wt{ght}^{(0)} C_{jh}^{(0)}=\sum_{g,h,t} \wt{ght}^{(0)}=\sum_{g,h}\left(\sum_{i,j,t} \At{ijt}\right)/K^2=\sum_{i,j,t} \At{ijt}$, which is the total number of edges in the network. Without this scaling, the initial parameters can have very small magnitudes, such that the mean total number of trips from the TDMM-SBM with these initial parameters is much smaller than the total number of trips in the data. In that situation, the initial likelihood is much smaller than the likelihood at the MLE.

To ensure that our estimated parameters are nonnegative, we use the following change of variables: $\exp(\tilde{\omega}^{(n)})=\omega^{(n)}$ and $\exp(\tilde{\mathbf{C}}^{(n)})=\mathbf{C}^{(n)}$. We can then write the gradient descent as 
\begin{align*}
	\tilde{\mathbf{\omega}}^{(n+1)} &= \tilde{\mathbf{\omega}}^{(n)}+\eta^{(n)}\nabla_{\mathbf{\omega}} \ell(\mathbf{C}^{
(n)},\mathbf{\omega}^{
(n)})\exp(\tilde{\mathbf{\omega}}^{
(n)})\,, \\
	\tilde{\mathbf{C}}^{(n+1)} &= \tilde{\mathbf{C}}^{(n)}+h^{(n)}\nabla_{\mathbf{C}}\ell(\mathbf{C}^{(n)},\mathbf{\omega}^{(n+1)})\exp(\tilde{\mathbf{C}}^{(n)})\,,
\end{align*}
where $h^{(n)}$ and $\eta^{(n)}$ are small positive numbers. From the definitions of $\tilde{C}^{(n)}$ and $\tilde{\omega}^{(n)}$, we write
\begin{align*}
	\mathbf{\omega}^{(n+1)} &= \mathbf{\omega}^{(n)}\exp(\eta^{(n)}\nabla_{\mathbf{\omega}} \ell(\mathbf{C}^{(n)},\mathbf{\omega}^{(n)}) \mathbf{\omega}^{(n)})\,,\\
\mathbf{C}^{(n+1)} &= \mathbf{C}^{(n)}\exp(h^{(n)}\nabla_{\mathbf{C}} \ell(\mathbf{C}^{(n)},\mathbf{\omega}^{(n+1)}) \mathbf{C}^{(n)})\,.
\end{align*}
We take the exponential of a vector to signify the application of the exponential to each component of the vector. Let $h^{(0)}=\eta^{(0)}=\Delta>0$ be the fixed initial step size.  For our application, we choose $\Delta =10^{-4}$. We generate two candidate updates for $\mathbf{\omega}^{(n+1)}$ for the first step in our algorithm using $h^{(n+1)}=1.2\,h^{(n)}$ and $h^{(n+1)}=0.8\,h^{(n)}$, and we choose the one that gives the $\mathbf{\omega}^{(n+1)}$ value that yields the larger $\ell(\mathbf{C}^{(n)},\mathbf{\omega}^{(n+1)})$. Similarly, we choose the one of $\eta^{(n+1)}=1.2\,\eta^{(n)}$ or $\eta^{(n+1)}=0.8\,\eta^{(n)}$ that gives the $\mathbf{C}^{(n+1)}$ value that yields the larger $\ell(\mathbf{C}^{(n+1)},\mathbf{\omega}^{(n+1)})$. 

We compute the gradient of the log-likelihood function $\ell$ using the chain rule. Recall that we compute the log-likelihood in two parts. One part is the computation of the mean $\mt{ijt}=\sum_{g,h}^K C_{ig}\omega_{gh}^t C_{jh}$ of the number of trips from node $i$ to node $j$ at time $t$. We then insert the expression for the mean into $\ell=\sum_t\sum_{i,j}\left(\At{ijt}\log(\mt{ijt})-\mt{ijt}\right)$. We compute the derivative of $\ell$ with respect to $\mt{ijt}$ to obtain
\begin{align*}
    \frac{\partial \ell}{\partial \mt{ijt}} = \frac{\At{ijt}}{\mt{ijt}} - 1\,.
\end{align*}
The derivative of $\mt{ijt}$ with respect to $C_{kg}$ is 
\begin{align*}
    \frac{\partial \mt{ijt}}{\partial C_{kg}} =&
        \delta_{ki}\sum_{h}
            \wt{ght} C_{jh} + \delta_{kj} \sum_{h}C_{ih} \wt{hgt} \,,
\end{align*}
where $\delta_{ab}$ denotes the Kronecker delta (i.e., $\delta_{ab}=1$ if $a=b$ and $\delta_{ab}=0$ if $a \neq b$). The derivative of $\mt{ijt}$ with respect to $\omega_{ght}$ is 
\begin{align*}
	\frac{\partial \mt{ijt}}{\partial\wt{ght}}=C_{ig}C_{jh}\,.
\end{align*}
 Using the above calculations, we see that the derivatives of $\ell$ with respect to $C_{kg}$ and $\omega_{ght}$ are
\begin{align*}
	\frac{\partial \ell} {\partial C_{kg}} 
&=\sum_{t=0}^{23}\sum_{i,j\in\mathcal{N}}\frac{\partial \ell}{\partial \mt{ijt}}\frac{\partial \mt{ijt}}{\partial C_{kg}}\\
&=
\sum_{t=0}^{23}
\left(\sum_{j\in\mathcal{N}}\left(\frac{\At{kjt}}{\mt{kjt}}-1\right) \sum_{h}\wt{ght}C_{jh}+
\sum_{i\in\mathcal{N}}\left(\frac{\At{ikt}}{\mt{ikt}} -1\right) \sum_{h}C_{ih}\wt{hgt}\right) \,,
\\
    \frac{\partial\ell}{\partial \wt{ght}} &= \sum_{i,j\in\mathcal{N}}\frac{\partial \ell}{\partial \mt{ijt}}\frac{\partial \mt{ijt}}{\partial \wt{ght}}
       \\&= \sum_{i,j\in\mathcal{N}} 
            \left(\frac{\At{ijt}}{  \mt{ijt}}-1\right) 
            C_{ig} C_{jh}\,.
\end{align*}

We run gradient descent until four significant digits of the base-10 floating-point representation of the log-likelihood \eqref{logroll} do not change for 600 consecutive steps. For the networks that we examine, this usually takes between 600 and 5000 iterations, with models with more blocks generally needing more iterations to reach this stopping criterion. Because of the non-convexity of the log-likelihood function \eqref{logroll}, we are not guaranteed to reach a global optimum. Most of the time, our method converges to an interesting local optimum (which may also be a global optimum) that appears to reveal functional roles of bicycle stations (see Section \ref{Results}). Our results produce recognizable block-connectivity parameters $\omega_{ght}$ (e.g., home--work commute patterns and leisure-usage patterns) and the parameters $C_{ig}$ indicate known spatial divisions of the stations (e.g., residential versus commercial districts). In some cases, however, our algorithm converges to an uninteresting local optimum; for example, sometimes the block-assignment parameters $C_{ig}$ for each station appear as if they are assigned independently at random. To improve our results, we run our algorithm repeatedly (specifically, $10$ times for each network) and store the estimate with the largest likelihood. {We compare the parameters that we obtain from gradient descent to those that we obtain by running a Hamiltonian Monte Carlo (HMC) sampling method in a Bayesian framework with weak priors. (We implement this sampling method in \textsf{Stan} \cite{stan2017jss, rstan2018}.) The log-likelihoods that result from our gradient-descent method are as good or better than those that we obtain with the HMC method. The HMC method is more computationally and memory intensive than our gradient-descent method, although it may be preferable in applications in which one has meaningful prior information about parameters.} Improving our optimization method and investigating trade-offs between accuracy and efficiency are worthwhile topics for future work. For instance, it likely will be beneficial to adapt optimization methods for related time-dependent SBMs \cite{xing2010dmsbm, yang2011dsbm, ho2011dmsbm, xu2014dsbm, matias2017semi} to the optimization of our SBMs.

The repository at {\url{https://github.com/jcarlen/tdsbm_supplementary_material}} has our \textsf{Python} implementation of our gradient-descent method and code for our inference in \textsf{R} using \textsf{Stan}.


\subsection{Inference using the TDD-SBM}

To fit our TDD-SBM, we use a Kernighan--Lin-type (KL-type) algorithm \cite{kernighanlin70} that we base on the one in \cite{karrer2011stochastic}. (In \cite{karrer2011stochastic}, Karrer and Newman noted that their KL-type algorithm is faster than node-switching algorithms and achieves results that are similar to those from such algorithms for their time-independent degree-corrected SBM.) Given $K$ blocks, we initialize the KL-type algorithm by assigning each node to a block uniformly at random, so each node has a probability of $\frac{1}{K}$ of being assigned to a given block. The algorithm then calculates the best possible block reassignment for any single node with respect to the associated change in log-likelihood (either the largest increase or the smallest decrease). Subsequently, we make the best reassignment for a different node, which we again choose uniformly at random, with respect to change in log-likelihood. The algorithm cycles through all nodes; a key feature of the algorithm is that a node that has been reassigned cannot move again until all other nodes have moved. One set of sequential reassignment of all nodes constitutes one step of the algorithm. The algorithm then searches all of the states (with respect to block membership of nodes) that have occurred during the step, and it selects the state with the maximum log-likelihood of any state during that step. This state is the starting point for the next step of the algorithm. A single run of the algorithm finishes when a step does not increase the log-likelihood beyond a preset tolerance value near $0$. (In practice, we use $1 \times 10^{-4}$.) To find block assignments that are as good as possible, we do many runs of the algorithm for each network; this process is easily parallelizable. In our examples, we use $50$ runs per network. We initialize each run randomly, as described above.  

Another key feature of the KL-type algorithm is that changes in the block membership of nodes affect only the terms of the objective function that involve the origin and destination blocks of the change. The simplified objective function \eqref{tdd-sbm-objective} is a sum over block-pair terms over $T$ time layers. Consequently, we do not need to recalculate the full objective function at each step. {This estimation method is practical for networks with up to a few thousand nodes, depending on available computing resources and on whether one can execute runs of the algorithm 
in parallel.

We implement our KL-type algorithm for TDD-SBM in \textsf{R} using \pkg{Rcpp} \cite{r18, eddelbuettel11rcpp}. The back-end calculations are in \textsf{C++} for speed, and we return results in \textsf{R} to enable visualization and other downstream tasks. Our implementation can also estimate time-independent SBMs, including directed and/or degree-corrected ones, {a variant of TDD-SBM variant without degree correction, and a TDD-SBM with directed degree-correction parameters $\theta_i^{\text{in}}$ and $\theta_i^{\text{out}}$}. This facilitates comparison of the results of inference from various related SBMs. See \url{https://github.com/jcarlen/sbm} for our \textsf{R} package \textbf{\texttt{sbmt}} for parameter estimation for the TDD-SBM. We {provide} code (which uses the \textbf{\texttt{sbmt}} package) at \url{https://github.com/jcarlen/tdsbm_supplementary_material} to replicate our examples in Section \ref{Results}.


\section{Evaluation of Model Fitting for Synthetic Networks}\label{SimStudy}

In this section, we generate data using three network models --- our TDD-SBM, a variant of our TDD-SBM without degree correction, and our TDMM-SBM. For networks that we generate from discrete-membership models, we estimate model parameters using the TDD-SBM, the TDD-SBM variant without degree correction, and the Poisson-process SBM (PPSBM) of Matias et al.~\cite{matias2017semi}.\footnote{
{We also apply} our TDD-SBM to the London bicycle-sharing network that was studied by Matias et al.~in \cite{matias2017semi} in the code at \url{https://github.com/jcarlen/tdsbm_supplementary_material}.} (Their model also does not have degree correction.) For the networks that we generate from the TDMM-SBM, we estimate model parameters using the TDMM-SBM and the TDD-SBM.} We compare how well these estimated models fit the generated data. We illustrate three key points. First, our estimation methods are able to recover data-generating parameters for generated examples. Second, incorporating degree correction in our models facilitates the identification of blocks with similar temporal activity patterns, as opposed to ones with similar activity levels (i.e., similar degrees), when nodes have heterogeneous degrees. Third, fitting our discrete-membership model (TDD-SBM) to data that is generated from our mixed-membership model (TDMM-SBM) can result in worse parameter estimates with respect to the likelihood of observed data {than when fitting to these data using our TDMM-SBM. That is, one can lose information when modeling mixed-membership networks as discrete-membership networks.}


We set up our generated networks as follows. We construct TDD-SBM and TDMM-SBM networks with 16 time layers to facilitate comparison to the PPSBM. In Figure \ref{fig:sim_omega_3}, we show the mean edge weight between nodes in a given pair of blocks for each time layer in the TDD-SBM. We design the underlying curves to resemble patterns that we observed when studying the urban bicycle-sharing networks. {Using these mean edge weights and a given number of nodes in each block, we calculate the expected sums of the edge weights from one block to another that originate in a given layer (the MLE of $\mathbf{\omega}$ in \eqref{w_ght-MLE}). These are the block-connectivity parameters $\mathbf{\omega} = \{\omega_{ght}\}$ that we use to generate our data.} 

We construct both two-block and three-block networks, where the former only uses values in the upper $2 \times 2$ panel of Figure \ref{fig:sim_omega_3}. For simplicity, we divide block membership equally between blocks for the TDD-SBM. For example, if there are 30 nodes and two blocks, 15 nodes belong to each block. We assume that {the degree-correction parameters $\mathbf{\theta}$ take one of five distinct values} $\{1x, 2x, \ldots, 5x\}$, which are distributed evenly within each block; the value of $x$ satisfies the blockwise sum constraints on $\mathbf{\theta}$.

\begin{figure}[H]
\flushleft
\includegraphics[scale=.66]{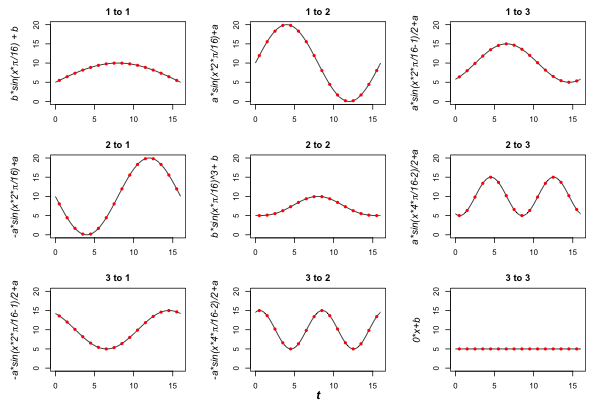}
\caption{In the TDD-SBM networks that we construct, the red dots indicate the mean weights for an edge from block $g$ to block $h$ at time $t$, where $g,h \in \{1,2,3\}$ and $t \in \{0.5,1.5,\ldots,15.5\}$. The title of each panel indicates the blocks $g$ and $h$ (e.g., `1 to 2').}
\label{fig:sim_omega_3}
\end{figure}

With these parameters and conditions, we construct multilayer networks using our TDD-SBM and TDMM-SBM. In each case, we produce $\nsims$ instantiations of the networks. All of the networks that we generate are directed, and we allow self-edges. We also generate and fit networks using a version of our TDD-SBM without degree correction. Discussion of this model is beyond the scope of this paper, but we implemented parameter estimation for it in our \textbf{\texttt{sbmt}} package. Its log-likelihood is 
\begin{align*}
	\ell(G;\mathbf{\omega}) = 
	\sum_{t}\sum_{g,h}\sum_{i\in g, j\in h} \left[\At{ijt}\log(\wt{ght}) - \wt{ght} - \log\left(A_{ijt}!\right) \right]\,,
\end{align*}
which is the log-likelihood of the TDD-SBM without the degree-correction terms $\mathbf{\theta}$.

For the networks that we generate (using the TDD-SBM or the TDD-SBM without degree correction), we estimate and compare their model parameters by fitting them using the TDD-SBM, the TDD-SBM without degree correction, and the PPSBM of Matias et al. \cite{matias2017semi} (which is implemented in the \textbf{\texttt{ppsbm}} package in \textsf{R} \cite{giorgi2018ppsbm}). PPSBM is a time-dependent SBM model that is similar to TDD-SBM without degree correction. It differs in that it assumes continuous-time networks and either adaptively splits the data into equal-sized bins (to use an approach based on non-parametric histograms) or uses a kernel-intensity estimator of parameters over time. For our comparison, we implement the former approach because the data input structure aligns well with our preprocessed data. We estimate the parameters of the PPSBM using a variational expectation-maximization (EM) algorithm. The \textbf{\texttt{ppsbm}} package includes a method to select the number of blocks using integrated classification likelihood (ICL).

In Table \ref{table:tdd_sim_results}, we show the results of networks that we generate using the TDD-SBM and the TDD-SBM without degree correction. The first two columns describe the number of blocks ($K$) and number of nodes ($N$) that we use to construct networks. (Recall  {that we pre-determine these values} in our models, so they are given.) We generate these networks using our TDD-SBM with and without degree correction as indicated in the `Gen DC' column. We indicate the model type that we use to fit the data in the `Fit Method' column, and we indicate whether it incorporates degree correction in the `Fit DC' column.

\begin{table}[H]
\caption{Evaluation of parameter-estimation methods for data that we generate using the TDD-SBM and the TDD-SBM without degree correction. We estimate parameters using both of these models and also using the PPSBM of Matias et al.~\cite{matias2017semi}. For each experiment, we indicate the number of blocks ($K$) and the number of nodes ($N$) in the networks. The `Gen DC' column indicates whether or not we use degree correction when generating the data, and the `Fit DC' column indicates whether or not we use degree correction when we fit the networks. The `Fit Method' column indicates which model we use for fitting. To the right of the vertical line, we show the mean values over $\nsims$ instantiations of the networks. We show the associated standard deviations in parentheses. The `ARI' column gives the adjusted Rand index between the inferred block assignments and the ground-truth assignments. The `MAPE' column gives the mean absolute-percentage error, expressed as a fraction, between the block-connectivity parameters $\omega$ and $\hat{\omega}$. In the `Gen LLIK' column, we give the unnormalized log-likelihoods of the generated networks under the data-generating model. In the `Diff LLIK' column, we give the result of subtracting the unnormalized log-likelihoods of the generated networks using the data-generating model from the unnormalized log-likelihoods using the estimated model.}
\label{table:tdd_sim_results}
\centering
\scalebox{0.7}{
\begin{tabular}{lllll|lllll}
  \hline
 $K$ & $N$ & Gen DC & Fit DC & Fit Method & ARI & MAPE (for $\hat{\mathbf{\omega}}$) & Gen LLIK & Diff LLIK \\ 
  \hline
  2 & 30 & F & F & TDD-SBM & 1 (0) & 0.03 (0) & 174414 (713) & 32 (6) \\ 
  3 & 30 & F & F & TDD-SBM & 1 (0) & 0.03 (0) & 169917 (784) & 70 (9) \\ 
  2 & 30 & T & F & TDD-SBM & 1 (0) & 0.03 (0) & 204808 (952) & $-$30390 (221) \\ 
  3 & 30 & T & F & TDD-SBM & 0.67 (0) & 1.8 (0.02) & 200581 (878) & $-$16673 (151) \\ 
  2 & 30 & F & T & TDD-SBM & 1 (0) & 0.03 (0) & 174467 (861) & 45 (7) \\ 
  3 & 30 & F & T & TDD-SBM & 1 (0) & 0.03 (0) & 169776 (855) & 86 (9) \\ 
  2 & 30 & T & T & TDD-SBM & 1 (0) & 0.03 (0) & 204964 (941) & 32 (6) \\ 
  3 & 30 & T & T & TDD-SBM & 1 (0) & 0.03 (0) & 200687 (1022) & 71 (8) \\ 
  2 & 30 & F & F & PPSBM & 1 (0) & 0.03 (0) & 174475 (827) & 31 (5) \\ 
  3 & 30 & F & F & PPSBM & 1 (0) & 0.03 (0) & 169928 (983) & 59 (11) \\ 
  2 & 30 & T & F & PPSBM & 0.04 (0.06) & 2.71 (0.36) & 204907 (887) & $-$31342 (310) \\ 
  3 & 30 & T & F & PPSBM & 0.16 (0.04) & 1.05 (0.22) & 200551 (899) & $-$18325 (951) \\ 
  \hline
  2 & 90 & F & F & TDD-SBM & 1 (0) & 0.01 (0) & 1569720 (2702) & 33 (6) \\ 
  3 & 90 & F & F & TDD-SBM & 1 (0) & 0.01 (0) & 1529181 (2512) & 72 (9) \\ 
  2 & 90 & T & F & TDD-SBM & 1 (0.04) & 0.04 (0.29) & 1843773 (2548) & $-$273917 (993) \\ 
  3 & 90 & T & F & TDD-SBM & 0.7 (0) & 1.8 (0.01) & 1805703 (2600) & $-$150714 (479) \\ 
  2 & 90 & F & T & TDD-SBM & 1 (0) & 0.01 (0) & 1569907 (2851) & 76 (8) \\ 
  3 & 90 & F & T & TDD-SBM & 1 (0) & 0.01 (0) & 1528897 (2574) & 115 (10) \\ 
  2 & 90 & T & T & TDD-SBM & 1 (0) & 0.01 (0) & 1843215 (2921) & 32 (5) \\ 
  3 & 90 & T & T & TDD-SBM & 1 (0) & 0.01 (0) & 1806107 (3201) & 74 (9) \\ 
  2 & 90 & F & F & PPSBM & 1 (0) & 0.01 (0) & 1569564 (2489) & 32 (5) \\ 
  3 & 90 & F & F & PPSBM & 1 (0) & 0.01 (0) & 1528835 (2333) & 63 (9) \\ 
  2 & 90 & T & F & PPSBM & 0.01 (0) & 2.77 (0.01) & 1843826 (3099) & $-$281730 (755) \\ 
  3 & 90 & T & F & PPSBM & 0.2 (0) & 1.1 (0.1) & 1806072 (2952) & $-$167642 (3864) \\ 
   \hline
\end{tabular}
}
\end{table}

We evaluate the success of each estimation method at detecting the true blocks assignments by calculating the adjusted Rand index (ARI) using the implementation in the \textbf{\texttt{fossil}} package in \textsf{R} \cite{vavrek2011fossil}. We report these values in the `ARI' column of Table \ref{table:tdd_sim_results}. The ARI is a function of two group-membership vectors and achieves its maximum value of $1$ when they match perfectly (up to label permutations). We expect an ARI of $0$ when we compare two uniformly random permutations of block assignments that preserve the block sizes. 

We evaluate the ability of the estimation methods to recover the true value of the parameter $\mathbf{\omega}$ using the mean absolute-percentage error (MAPE) between $\hat{\mathbf{\omega}}$ and $\mathbf{\omega}$. The formula for the MAPE is $\text{MAPE}=\frac{1}{ght}\sum_{ght}|\frac{\hat{\omega}_{ght} - \omega_{ght}}{\omega_{ght}}|$, which we report in the `MAPE' column of Table \ref{table:tdd_sim_results}. We also compare the unnormalized log-likelihoods of the generated networks under the data-generating model (in the `Gen LLIK' column) to the unnormalized log-likelihoods under the estimated model. Given the similarities of the candidate models, this value provides an overall measure of how well we estimate the Poisson means of the edge weights. We show the differences between these two quantities in the `Diff LLIK' column, where a positive value signifies that the data is more likely under the estimated model.

We find perfect matches (i.e., an ARI of $1$ with standard deviation of $0$) between the true and estimated block-membership assignments in all cases when the data-generation and estimation methods match and when the estimation model (TDD-SBM) is a generalization of the network-generation model (TDD-SBM without degree correction). We also find perfect matches when the data is generated from the TDD-SBM without degree correction and is estimated by the PPSBM; this illustrates the similarity of these two models. In all of those cases, there is very little error in $\hat{\mathbf{\omega}}$, as indicated by the small MAPE values, and the likelihood of the generated data under the estimated model is slightly larger than under the data-generating model. (Recall that we fit the former to the generated data.) This indicates that our TDD-SBM estimation method is effective in these examples. 

When we generate networks using the TDD-SBM with degree correction, the two methods without degree correction (i.e., the TDD-SBM variant and PPSBM) do a poor job of fitting the data. In the three-block networks that we generate, we obtain large values of both ARI and MAPE. Even when we recover the blocks as well as in the two-block examples, we obtain much smaller log-likelihoods of the generated networks under the estimated models than under the data-generating model. This is the case because using a model without degree correction implies that all edges between a given pair of blocks are identically distributed, regardless of the nodes to which the edges are attached. This does not accurately reflect our generated data, which has heterogeneous degrees within blocks.

For the PPSBM estimation method, the results in Table \ref{table:tdd_sim_results} reflect parameter estimates when we set the number of blocks to be equal to the value (which is two or three in our examples) in our data-generation mechanism. When we apply the block-selection method of Matias et al.~\cite{matias2017semi} to models with between one and four blocks (using the implementation in the \textbf{\texttt{ppsbm}} package), we identify the true number of blocks when we generate the data without degree-correction. However, when we apply it to data that we generate using the TDD-SBM with degree correction and let it search for up to four blocks, we overestimate the number of blocks. Interestingly, the PPSBM method attempts to classify nodes based both on their roles and on degree-correction factors. SBMs without degree correction determine roles based both on overall degree and on connectivity patterns.

We are interested in how successful our mixed-membership model (i.e., TDMM-SBM) is able to recover data-generating parameters and how much log-likelihood we lose when we use a TDD-SBM to fit a network that we generate using the TDMM-SBM with degree correction. For each combination of number of blocks ($K$) and number of nodes ($N$), we draw values of $\mathbf{C}$ uniformly at random from $\{0, 1, \ldots, 5\}$ and then normalize each block so that $\sum_iC_{ig}=1$. We set the values of $\mathbf{\omega}$ equal to those that we used above for the degree-corrected TDD-SBM with corresponding numbers of blocks and nodes. We also include results for networks that we generate from parameters of TDMM-SBM that we determine from fitting to the Los Angeles bicycle-sharing network with two and three blocks. In those examples, $N=61$, because that is the number bicycle-docking stations in that network. We summarize our results in Table \ref{table:tdmm_sim_results}. 

Because block membership is mixed, we do not use ARI to compare $\hat{\mathbf{C}}$ with $\mathbf{C}$. Instead, we calculate the blockwise absolute error (BAE), which is the sum of absolute error averaged over the number of blocks. That is, $\text{BAE} = \frac{1}{K}\sum_{i,g}|\hat{C}_{ig} - C_{ig}|$. To evaluate the error of $\hat{\mathbf{\omega}}$, we calculate the MAPE. We also calculate a `pairwise' MAPE of $\hat{\mathbf{\omega}}$ (which we denote by MAPE$_p$ in Table \ref{table:tdmm_sim_results}), in which we average the absolute percentage error over each pair of blocks. That is, $\text{MAPE}_p = \frac{1}{K^2}\sum_{g,h}\frac{\sum_t|\hat{\omega}_{ght} - \omega_{ght}|}{\sum_t\omega_{ght}}$. This quantity is less susceptible than MAPE to {having very large values when individual values of $\omega_{ght}$ are near $0$.}

We compare the log-likelihoods of the generated networks under the data-generating TDMM-SBM (`Gen LLIK') to their log-likelihoods under the estimated parameters, and we report the difference (the latter minus the former) in the `Diff LLIK' column of Table \ref{table:tdmm_sim_results}. Finally, we fit a degree-corrected TDD-SBM to the generated networks and calculate the log-likelihood of the networks under that model. We report this quantity minus the log-likelihood from the `Gen LLIK' column in the `Diff LLIK Discrete' column.

\begin{table}[H]
\caption{Evaluation of parameter-estimation methods for data that we generate using the TDMM-SBM. For each experiment, we indicate the number of blocks ($K$) and the number of nodes ($N$) in the networks. To the right of the vertical line, we show the mean values over $\nsims$ instantiations of the networks. We show the associated standard deviations in parentheses. The `BAE' column gives the blockwise absolute errors for the block-membership parameters. The `MAPE' column gives the mean absolute-percentage error{, expressed as a fraction,} between the block-connectivity parameters $\omega$ and $\hat{\omega}$. We show `--' in this column when the result is extremely large due to values of $\omega_{ght}$ near $0$. The `MAPE$_p$' column gives the pairwise MAPE {of} $\hat{\mathbf{\omega}}$. In the `Gen LLIK' column, we give the unnormalized log-likelihoods of the generated networks under the data-generating TDMM-SBM. In the `Diff LLIK' column, we give the result of subtracting `Gen LLIK' from the unnormalized log-likelihoods {of the generated networks} {using} the estimated TDMM-SBM. In the `Diff LLIK Discrete' column, we give the result of subtracting `Gen LLIK' from the unnormalized log-likelihoods of the generated networks using the degree-corrected TDD-SBM.}\label{table:tdmm_sim_results}
\centering
\scalebox{0.7}{
\begin{tabular}{ll|llllll}
  \hline
 $K$ & $N$ & BAE (for $\hat{\mathbf{C}}$) & MAPE (for $\hat{\mathbf{\omega}}$) & MAPE$_p$ (for $\hat{\mathbf{\omega}}$) & Gen LLIK & Diff LLIK & Diff LLIK Discrete \\ 
  \hline
  2 & 30 & 0.02 (0.004) & 0.1 (0.01) & 0.04 (0.01) & 198447 (969) & 58 (8) & $-$2243 (58) \\ 
  3 & 30 & 0.22 (0.054) & 0.74 (0.21) & 0.52 (0.1) & 168880 (729) & 94 (23) & $-$241 (25) \\ 
  2 & 90 & 0.02 (0.001) & 0.09 (0.01) & 0.02 (0.01) & 1651845 (2806) & 117 (11) & $-$22830 (195) \\ 
  3 & 90 & 0.21 (0.077) & 0.48 (0.19) & 0.39 (0.09) & 1581507 (2343) & 135 (109) & $-$4221 (82) \\ 
  \hline
  2 & 61 & 0.05 (0.004) & -- & 0.08 (0.01) & $-$37351 (233) & 105 (10) & $-$597 (38) \\ 
  3 & 61 & 0.06 (0.005) & -- & 0.29 (0.03) & $-$35214 (211) & 168 (11) & $-$1569 (54) \\ 
  \hline
\end{tabular}
}
\end{table}

We find that the likelihood of the generated data when we use the estimated parameters is slightly larger than the likelihood of the data when we use the data-generating parameters. (Because of  the  randomness in the networks that we generate, we expect that a network that we generate using a given model will have a larger likelihood when we use the parameters that we fit to that network than when we use the data-generating parameters.)} Additionally, the likelihood of the generated data under parameters that we estimate using the TDD-SBM is significantly smaller than those that we obtain under the data-generating TDMM-SBM parameters, especially for the three-block models. (See the `Diff LLIK Discrete' column of Table \ref{table:tdmm_sim_results}.) In short, we gain a lot of information by using our mixed-membership model in this case.

The BAE, MAPE, and MAPE$_p$ values (i.e., the `error values') tend to be small for the two-block models, indicating that we are recovering the data-generating parameters fairly well in these cases. (We do not report MAPE values for generated networks when we use the estimated parameters of TDMM-SBM of Los Angeles, which has $N=61$ bicycle-sharing stations, because a few values of $\mathbf{\omega}$ that are very close to $0$ drastically inflate those values.) However, the values of BAE, MAPE, and MAPE$_p$ are much larger for the three-block models (except for the BAE for Los Angeles). The fact that the estimated parameters achieve similar likelihoods as the data-generating parameters despite fairly large values of BAE, MAPE, and MAPE$_p$ suggests that the TDMM-SBM is not completely identifiable in these cases. Although the BAE error values are not as large for the networks that we generate using parameter values that we estimate from the data from Los Angeles. it is important to be aware of this identifiability limitation when interpreting the parameters of our TDMM-SBM. This is especially the case for $\mathbf{\omega}$, given the large MAPE and MAPE$_p$ values of $\hat{\mathbf{\omega}}$ that we obtained for all three-block TDMM-SBMs.

The identifiability of time-dependent SBMs is a relatively unexplored research topic. Matias and Miele \cite{matias2017dynsbm} pointed out that the time-dependent SBM of Xu and Hero \cite{xu2014dsbm} is not fully identifiable, and they introduced a demonstrably identifiable (up to label permutations) time-dependent SBM. They also established that identifiability is an inherent limitation of time-dependent SBMs in which block-membership parameters and block-connectivity parameters vary over time, unless one applies additional constraints or penalization. As we described in Section \ref{Model}, our models only allow the block-connectivity parameters to vary. Further research is necessary to establish the conditions for identifiability of our time-dependent SBMs.


\section{Results on Bicycle-Sharing Networks}\label{Results}

We apply our models to bicycle-sharing networks in Los Angeles, San Francisco, and New York City. (See Section \ref{Data} for full descriptions of these data sets.) The networks that we examine in downtown Los Angeles and San Francisco are relatively small; they had 61 stations and 35 stations, respectively, at the time that we collected our data. The stations in these networks are concentrated in downtown areas, where residential buildings and high-rise office buildings are interspersed. The New York City network is much larger than the other two networks. It includes about 600 stations, which encompass a range of commercial areas, residential neighborhoods, parks, and manufacturing areas.

Before discussing our results in detail, we highlight some of our key insights. We present results for two-block models for the Los Angeles (see Section \ref{la}) and San Francisco (see Section \ref{sf}) bicycle-sharing networks that illustrate the ability of our models to detect functional roles in the form of `home' and `work' roles of bicycle-sharing stations. These results facilitate the ability to compare bicycle-sharing systems across cities. We fit discrete-membership and mixed-membership models with two, three, and four blocks to each of our data sets; and we fit five-block and six-block models in several cases. We provide the output of these models and all code to recreate our results at \url{https://github.com/jcarlen/tdsbm_supplementary_material}. In most cases, we find that additional blocks do not illuminate new functional roles beyond the home and work roles. For example, using block models with additional blocks for Los Angeles and San Francisco result in small variations of these roles but increase the noise in the time variation of the block-connectivity parameters $\hat{\omega}_{ght}$. {In Section \ref{nyc}, we present five-block TDMM-SBM and TDD-SBM} fits of a subset of the New York City bicycle-sharing network that illustrates another role (a `park' block) that does not appear distinctly when we employ models with fewer blocks. A method to select the number of blocks is beyond the scope of our paper, but we discuss aspects of this model-selection problem in Appendix \ref{ModelSelection} and illustrate it by fitting mixed-membership models with up to 10 blocks on the Los Angeles bicycle-sharing network.


\subsection{Downtown Los Angeles} \label{la}

\begin{figure}[H]\center
\includegraphics[scale=.15]{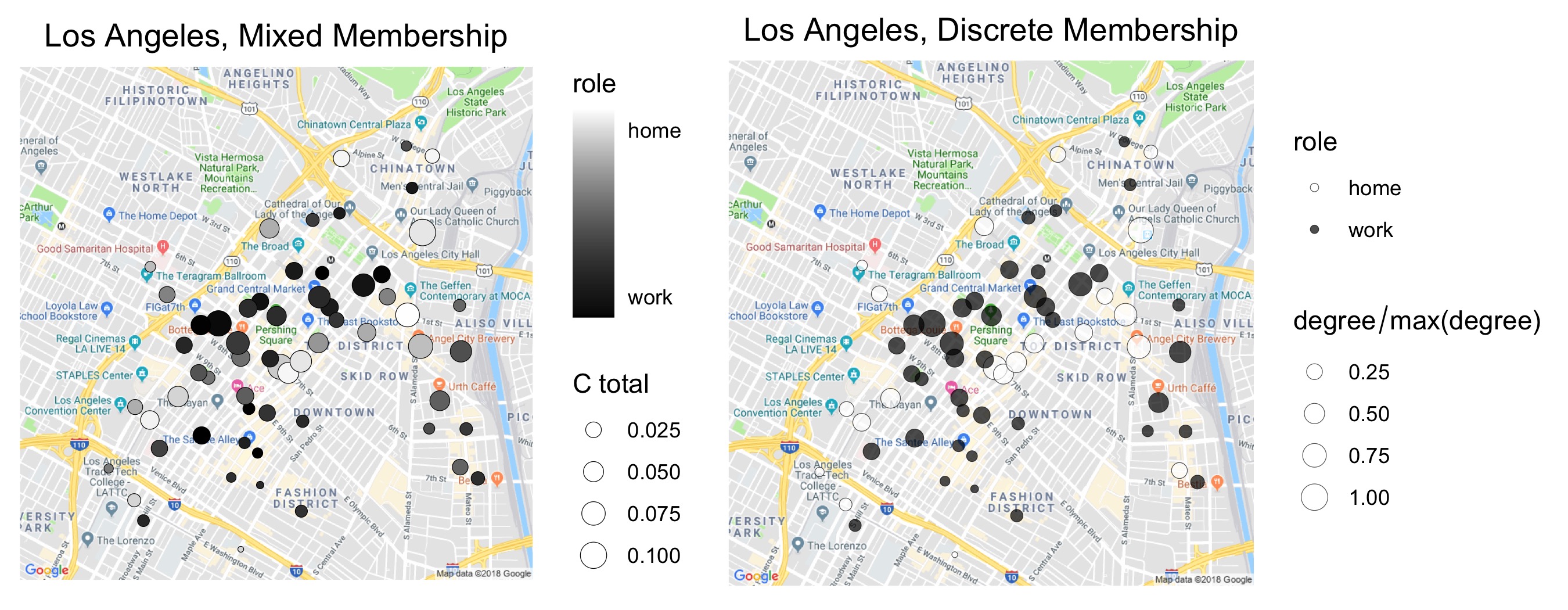}
\caption{Downtown Los Angeles bicycle stations classified using (left) a two-block TDMM-SBM and (right) a two-block degree-corrected TDD-SBM. The sizes of the nodes take continuous values. In the left panel, we scale the size of each node based on its value of $\sum_gC_{ig}$. In the right panel, we scale the size of each node based on its degree.}\label{fig:LA_mixed_discrete}
\end{figure}

In Figure \ref{fig:LA_mixed_discrete}, we show the mixed-membership (TDMM-SBM) and degree-corrected discrete (TDD-SBM) block assignments of two-block models of the downtown Los Angeles bicycle-sharing network. For the TDMM-SBM, we scale the size of each node $i$ in our plots based on its value of $\sum_gC_{ig}$. We refer to these sums as `C total' values. These values correlate strongly with node degree (specifically, the sum of the in-degree and the out-degree)}, which is evident in the similarity of node sizes in the left and right panels of Figure \ref{fig:LA_mixed_discrete}. For both models, we observe that home and work blocks are interspersed geographically. As part of our discussion in this subsection, we will describe our method for determining the block labels in Figure \ref{fig:LA_mixed_discrete}. 

The TDMM-SBM result reveals a group of stations (which we color in gray) in the left panel of Figure \ref{fig:LA_mixed_discrete} that are neither strongly home-identified nor strongly work-identified. Instead, they possess a roughly even mixture of the two types. For this network, the TDD-SBM output is very similar to what we obtain from a discretization of the TDMM-SBM output (where we discretize by assigning each node $i$ to the block with the maximum value of its $C_{ig}$ parameter), but this is not true for all of our bicycle-sharing networks.

Our model does not yield `home' and `work' labels for each block on its own, so we use the time-dependent block-connectivity parameter 
estimates $\hat{\omega}_{ght}$ to assign these labels. We assign the labels heuristically under the assumption that the `home' block is the origin of many trips to the work block in the morning and the `work' block is the origin of many trips to the home block in the evening. Figure \ref{fig:la_omega}, which shows $\hat{\omega}_{ght}$ for each possible value of $g$ and $h$, with the hour $t$ on the horizontal axis, supports our labeling. Given our labeling, we observe a clear peak in home-to-work traffic in the mornings and work-to-home traffic in the evenings. We make similar `home' and `work' assignments for San Francisco and New York City. In Los Angeles, the traffic within the work block peaks in the middle of the day. This may represent lunchtime errands, leisure activity, or tourist activity, as there are many tourist attractions in the downtown area. The traffic within the home block has a mild evening peak and has the least activity overall among the block pairs.

\begin{figure}[H] \center
\includegraphics[scale=.11]{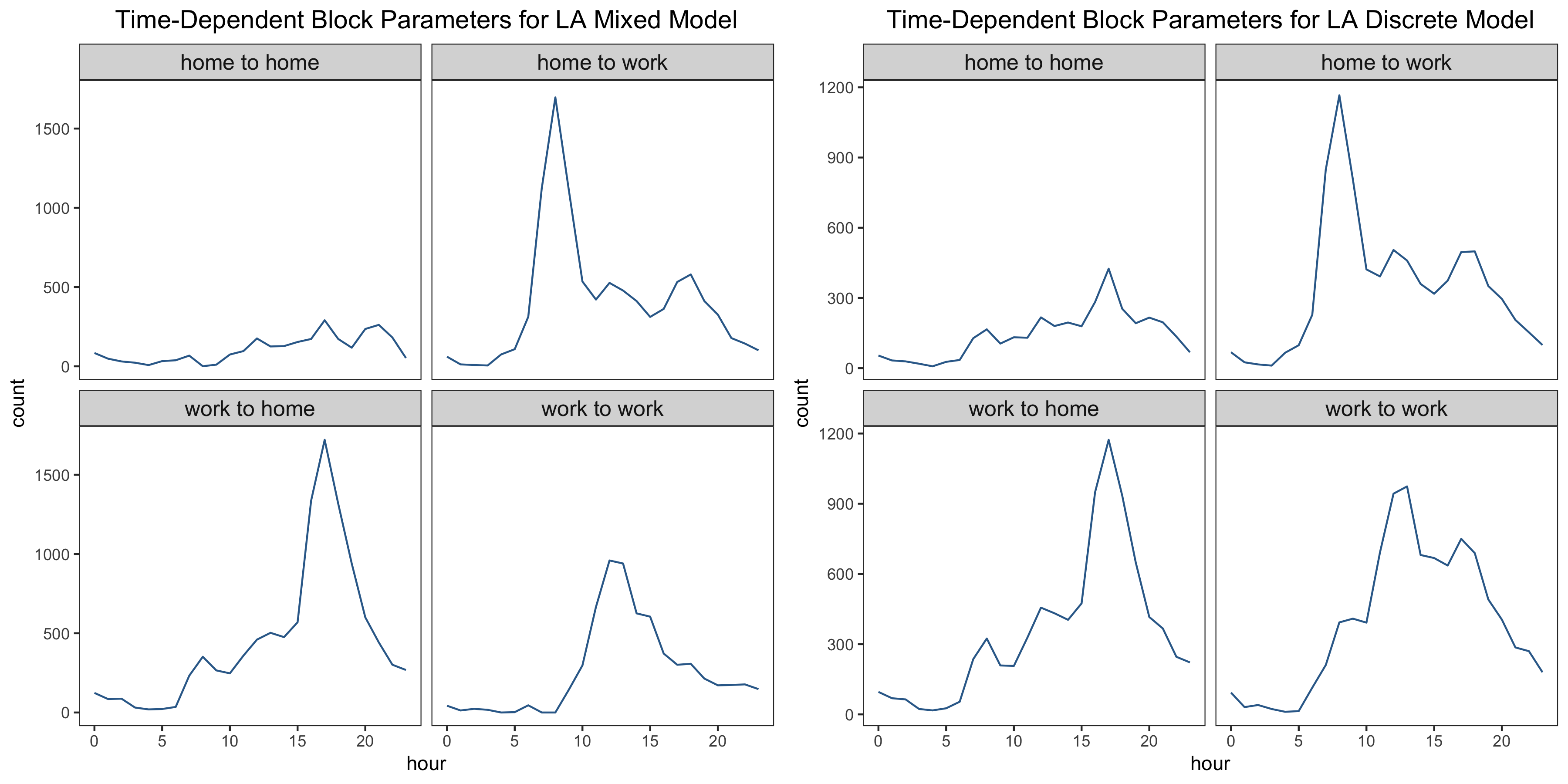}
\caption{Estimated time-dependent block-connectivity parameters $\hat{\omega}_{ght}$ for the two-block TDMM-SBM and the two-block TDD-SBM for downtown Los Angeles.}
\label{fig:la_omega}
\end{figure}

To further validate our block labels, we use the zoning map for downtown Los Angeles from the Los Angeles Department of City Planning \cite{la_zoning}.\footnote{Permission for use of these proprietary data is granted by the City of Los Angeles Department of City Planning. Copyright $\copyright$ 2015 City of Los Angeles. All Rights Reserved.} Zoning ordinances determine the allowable uses of city land. They distinguish land that is available for commercial uses, industrial uses, residential uses, park districts, and other uses. In the background of Figure \ref{fig:LA_Zoning}, we show a simplified version of the underlying zoning map; in this map, we have grouped similar designations). The industrial areas house a mixture of manufacturing and commercial uses. Public facilities include government buildings, public schools, parking under freeways, and police and fire stations \cite{la_zoning_dict}. In downtown Los Angeles, manufacturing and industrial areas are split cleanly from residential areas, whereas residential and commercial areas are intermixed across the bicycle-sharing system's coverage area.

\begin{figure}[H]\center
\includegraphics[scale=.15]{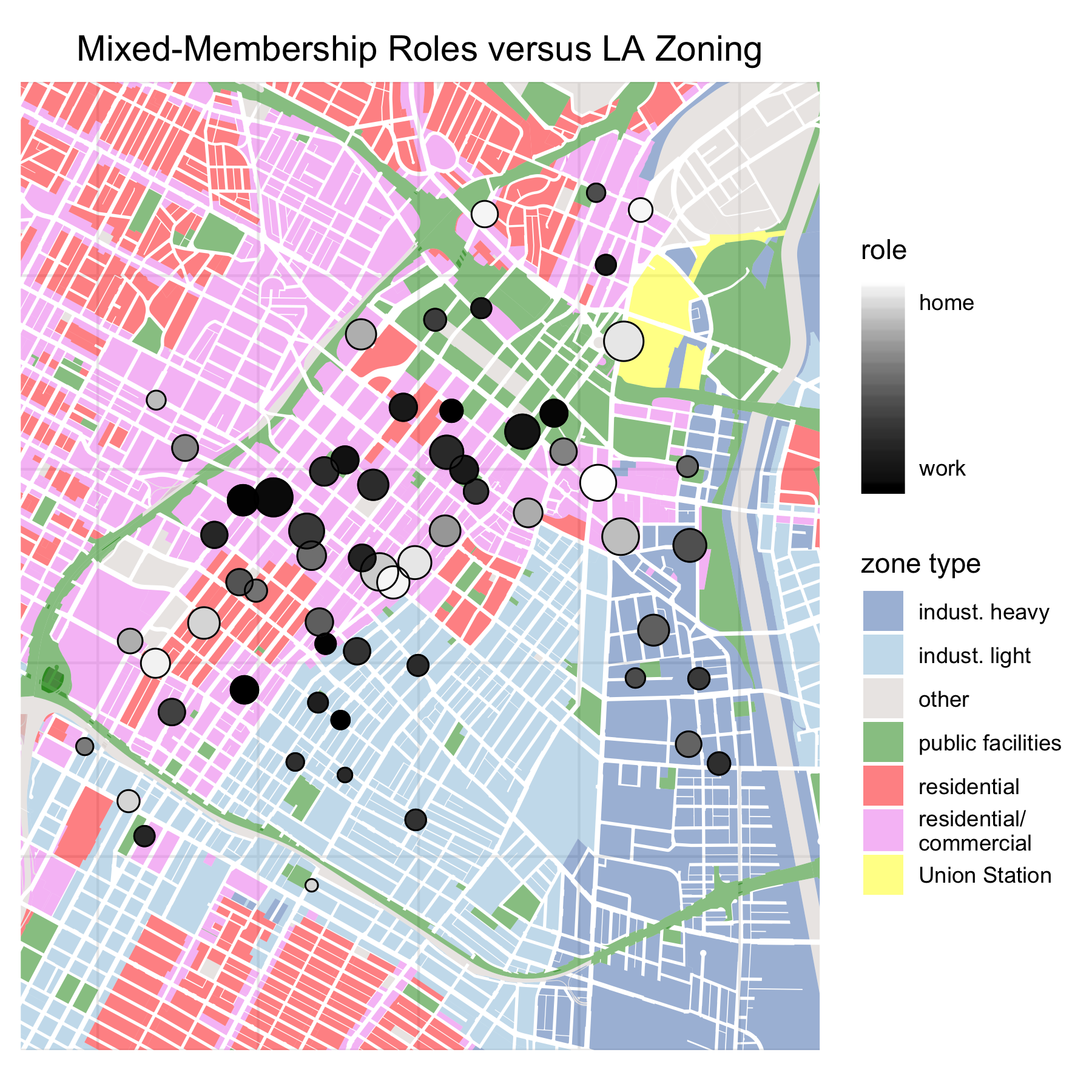}
\caption{Mixed-membership (TDMM-SBM) assignments of Los Angeles bicycle-sharing stations overlaid on a simplified LA zoning map. Industrial blocks include manufacturing and commercial areas. As in Figure \ref{fig:LA_mixed_discrete}, we scale the size of each node based on its value of $\sum_gC_{ig}$.}
\label{fig:LA_Zoning}
\end{figure}

Figure \ref{fig:LA_Zoning} illustrates that most stations that are strongly home-identified are in or near zones for pure residential use or mixed residential and commercial use. We find that many stations that are not predominantly home-identified or work-identified align with mixed-use commercial/residential zones. The discrete-role plot (see the right panel of Figure \ref{fig:LA_mixed_discrete}) has a stripe of `home' stations that cut diagonally through the `work' stations. In Figure \ref{fig:LA_Zoning}, we see that this aligns roughly with areas that are zoned for purely residential use. By contrast, industrial and public facility zones tend to host stations that are mostly work-identified, although some of the most strongly work-identified stations are in mixed-used areas.

One station that seems to deviate from the overall pattern is the heavily-trafficked station at Union Station. Although it is adjacent to a public facility zone with many government buildings, it is also strongly home-identified. This may seem surprising on its surface, but this classification is consistent with other home-identified stations because Union Station is a major transit hub for the Los Angeles metropolitan area. Accordingly, many morning trips originate there, as commuters transition from other forms of transportation, and many evening trips conclude there. This type of activity pattern that is sensibly associated with home-identified stations. Such idiosyncrasies of transit hubs also arise in our results for San Francisco and New York City.


\subsection{San Francisco}\label{sf}

In Figure \ref{fig:sf_cont_discrete}, we compare the two-block TDMM-SBM and two-block degree-corrected TDD-SBM for San Francisco. As we saw for Los Angeles, the San Francisco blocks are interspersed geographically and stations vary from strongly home-identified ones to strongly work-identified ones. The {two} most strongly home-identified station{s are located at} a major transit hub, the San Francisco Caltrain Station on 4th Street.

\begin{figure}[H]
\includegraphics[scale=.15]{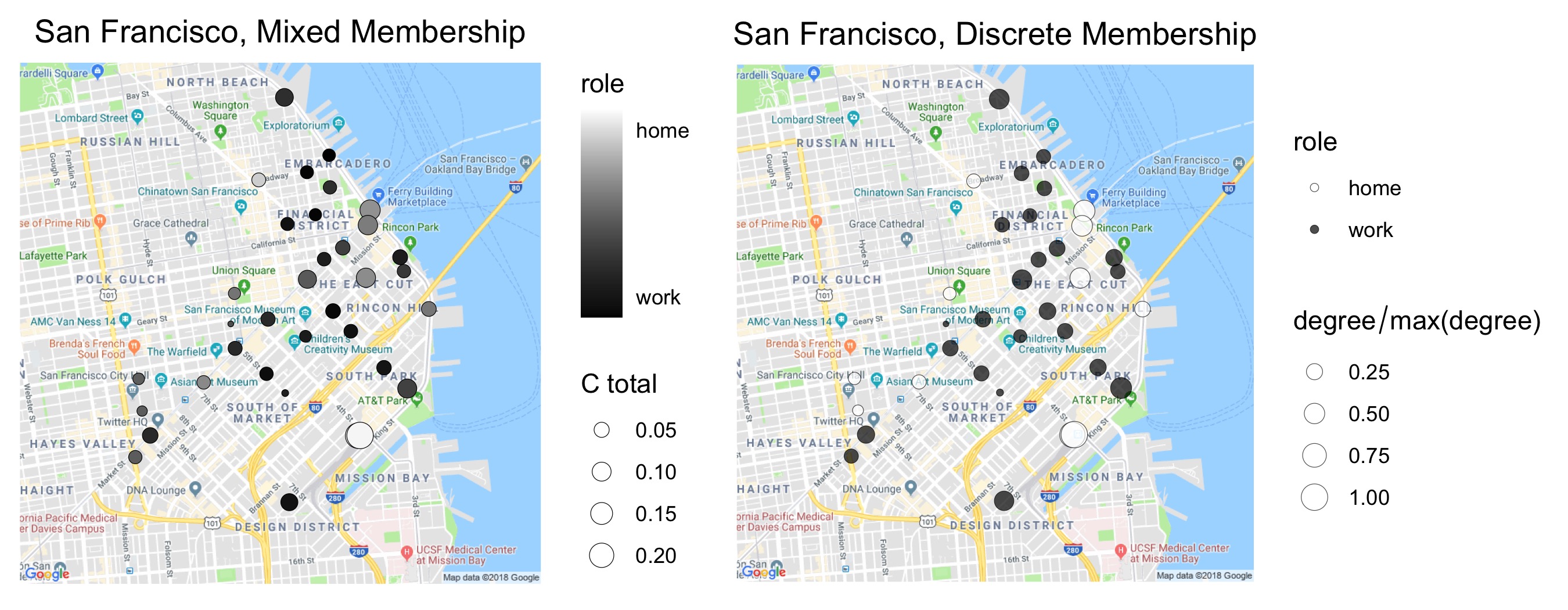}
\caption{San Francisco bicycle stations classified using (left) a two-block TDMM-SBM and (right) a two-block TDD-SBM. The sizes of the nodes take continuous values. In the left panel, we scale the size of each node based on its value of $\sum_gC_{ig}$. In the right panel, we scale {the size of} each node based on its degree.}
\label{fig:sf_cont_discrete}\label{fig:SF_mixed_discrete}
\end{figure}

In Figure \ref{fig:sf_omega}, we show the estimated traffic between the `home' and `work' blocks for the TDMM-SBM and the TDD-SBM. As in the downtown Los Angeles network, we observe inter-block commuting. However, unlike in downtown LA, the TDD-SBM reveals intra-block morning and evening peaks in both the home and work blocks. This may be due to last-mile commuting, such as using bicycle-sharing facilities to get to or from a train station. {We find that the ten most-popular trips within blocks (i.e., home-to-home travel or work-to-work travel when there are two blocks) that originate during the heavy commuting hours of 7--9 am and 4--6 pm all begin at major public transportation stops in the morning and end at those stops in the evening.} Recognizing last-mile usage is important for integrating bicycle sharing with nearby public transportation. One possible reason that we do not observe a similar phenomenon in downtown LA is that San Franciscans are more likely than residents of Los Angeles to use public transportation \cite{pubtranLAT2015}. The intra-block morning and evening peaks may also arise from the intermixing of commercial and residential uses of land. In other words, some travel within blocks may also constitute commuting.

\begin{figure}[H] \center
\includegraphics[scale=.11]{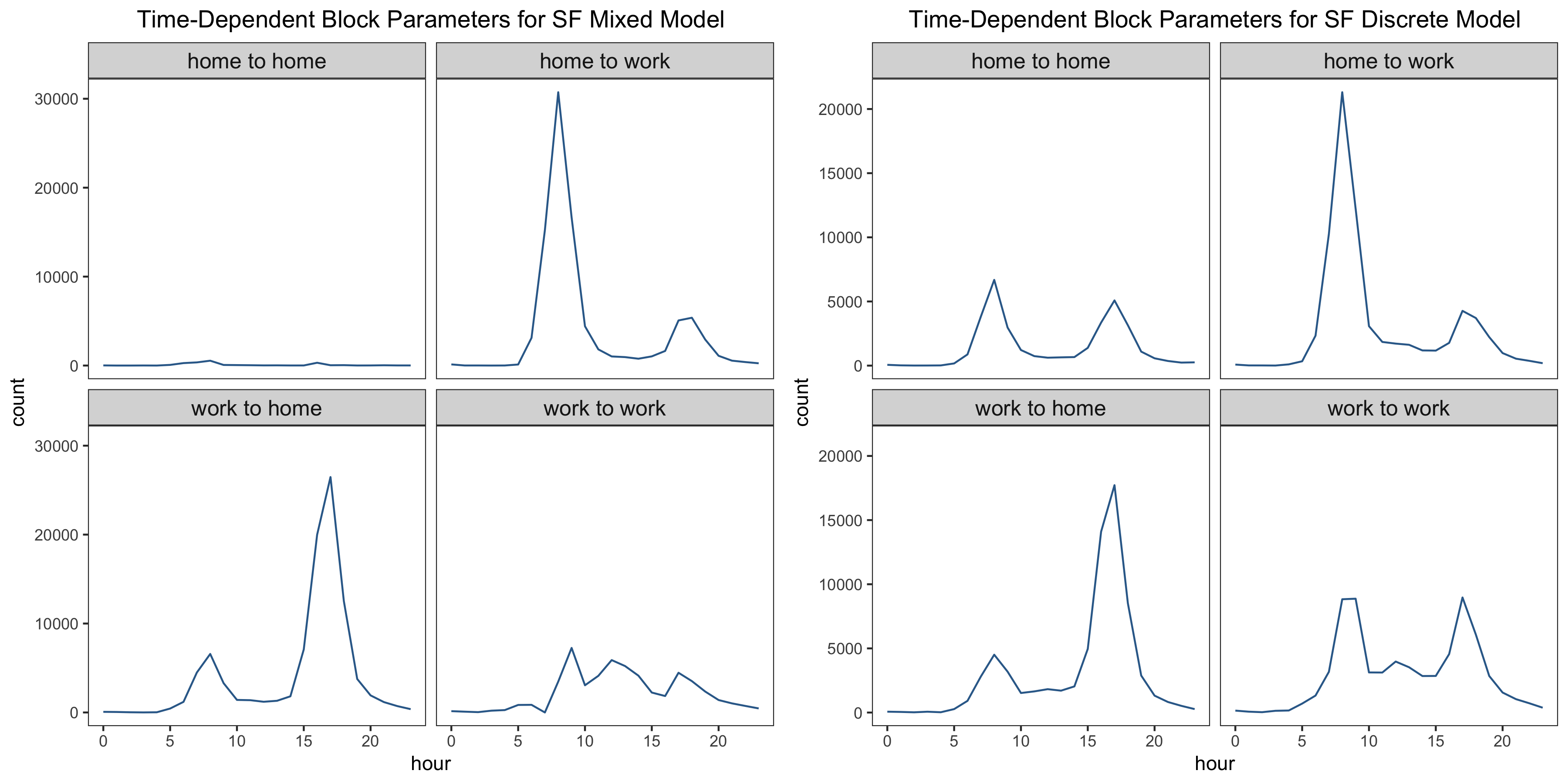}
\caption{Estimated time-dependent block-connectivity parameters $\hat{\omega}_{ght}$ for the two-block TDMM-SBM and two-block degree-corrected TDD-SBM for San Francisco.}
\label{fig:sf_omega}
\end{figure}


\subsubsection{Comparison to SBM Variations}

We briefly compare our above results for Los Angeles and San Francisco to {results that we obtain for these networks by employing a time-independent SBM and a time-dependent SBM without degree correction. We do this to illustrate that both time-dependence and degree correction are necessary to achieve our goal of detecting functional roles of bicycle-sharing stations.

To fit the time-independent SBM, we calculate the time-independent adjacency matrix $\tilde{A}_{ij}=\sum_{t=0}^{23}\At{ijt}$ and fit the two-block degree-corrected TDD-SBM to the associated time-aggregated network with a single layer. For the downtown Los Angeles network, we observe a clear geographically-based division in the results of the time-independent SBM (see Figure~\ref{fig:static_discrete}). For the San Francisco network, the differences between the blocks of the time-dependent SBM and time-independent SBM are less noticeable, although they are still present. This confirms that our time-dependent SBMs detect behavior that is not evident in the time-aggregated data.

\begin{figure}[H]\center
\includegraphics[scale=.15]{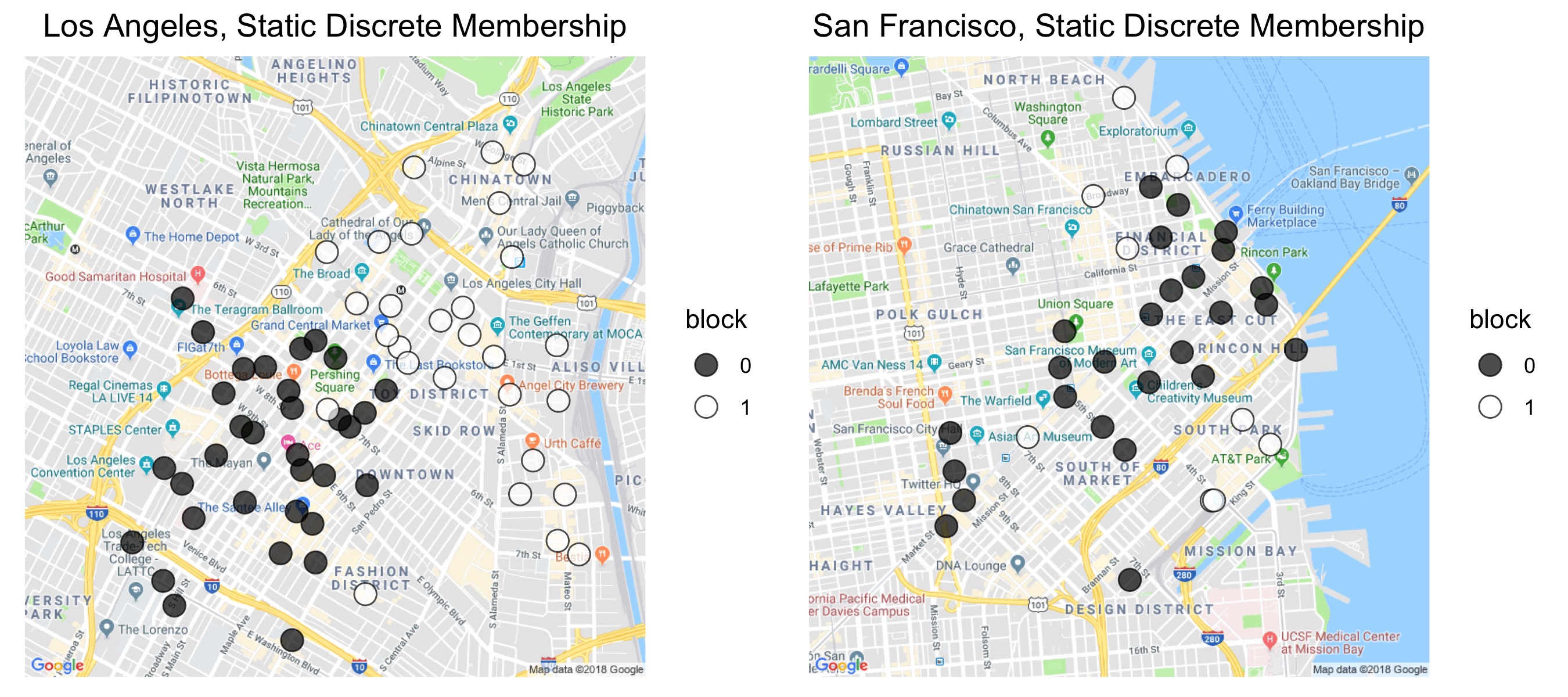}
\caption{Estimated blocks from our fit of a two-block degree-corrected TDD-SBM to time-aggregated bicycle-sharing networks in Los Angeles and San Francisco.}
\label{fig:static_discrete}
\end{figure}

We now fit the variant of our two-block TDD-SBM without degree correction (see Section \ref{SimStudy}) to the Los Angeles and San Francisco bicycle-sharing networks. In Figure~\ref{fig:nodc_discrete}, we see in this case that stations are now divided into one group of small-degree stations and one group of large-degree stations, where `degree' in the above sentence is the sum of the in-degree and the out-degree. These blocks differ significantly from those that we obtained using the TDD-SBM with degree correction. (See Figures \ref{fig:LA_mixed_discrete} and \ref{fig:sf_cont_discrete}.) For example, the TDD-SBM for Los Angeles with degree correction classifies the the high-traffic node at Union Station as a `home' station because of its frequent departures in the morning and frequent arrivals in the evening, whereas the the TDD-SBM without degree correction groups Union Station with other high-traffic stations that are predominantly `work' stations in the degree-corrected model.

\begin{figure}[H]\center
\includegraphics[scale=.15]{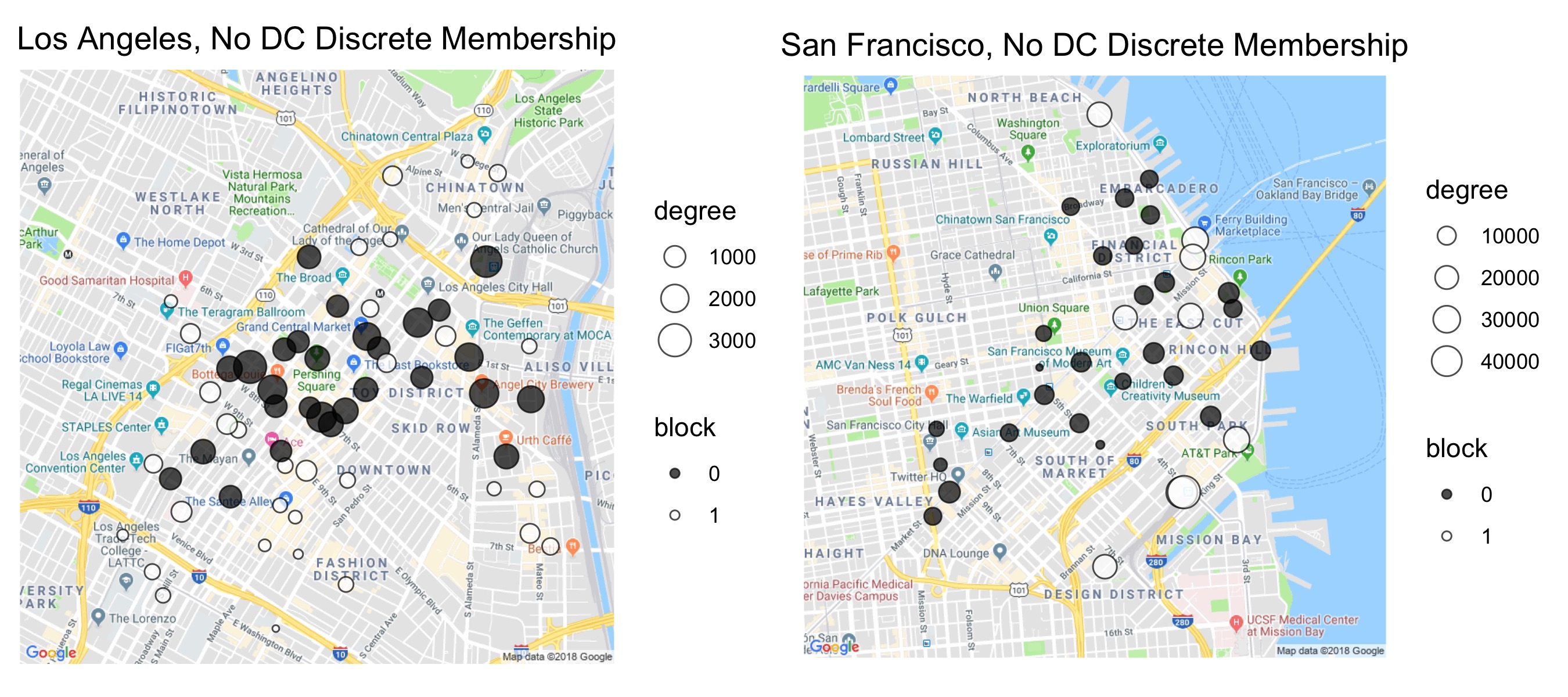}
\caption{Estimated blocks from a TDD-SBM variant without degree correction for bicycle-sharing networks in (left) Los Angeles and (right) San Francisco. We scale the size of each node based on its degree.}
\label{fig:nodc_discrete}
\end{figure}



\subsection{New York City}\label{nyc}

We fit the TDMM-SBM and the degree-corrected TDD-SBM to the New York City bicycle-sharing network with two through six blocks. Our results suggest that there is a limit to the size of networks for which our time-dependent SBMs can recover blocks that are based on functional roles of stations (rather than on geographical locations). Therefore, in the present subsection, we apply our models to a subset of the New York City bicycle-sharing network to gain insight about the functional roles of stations in the subset. We select this subset based on the results of applying our three-block TDD-SBM to the entire New York City network. We examine the subnetwork that is induced by the stations that are in one of these three blocks. We explore three-block models of the entire New York City network in Appendix \ref{appendix:nyc}. In those results, we infer that stations are assigned to blocks along geographical lines, rather than to blocks with distinct functions. 


\subsubsection{Manhattan subnetwork}

\begin{figure}[H]\center 
\includegraphics[scale=.11]{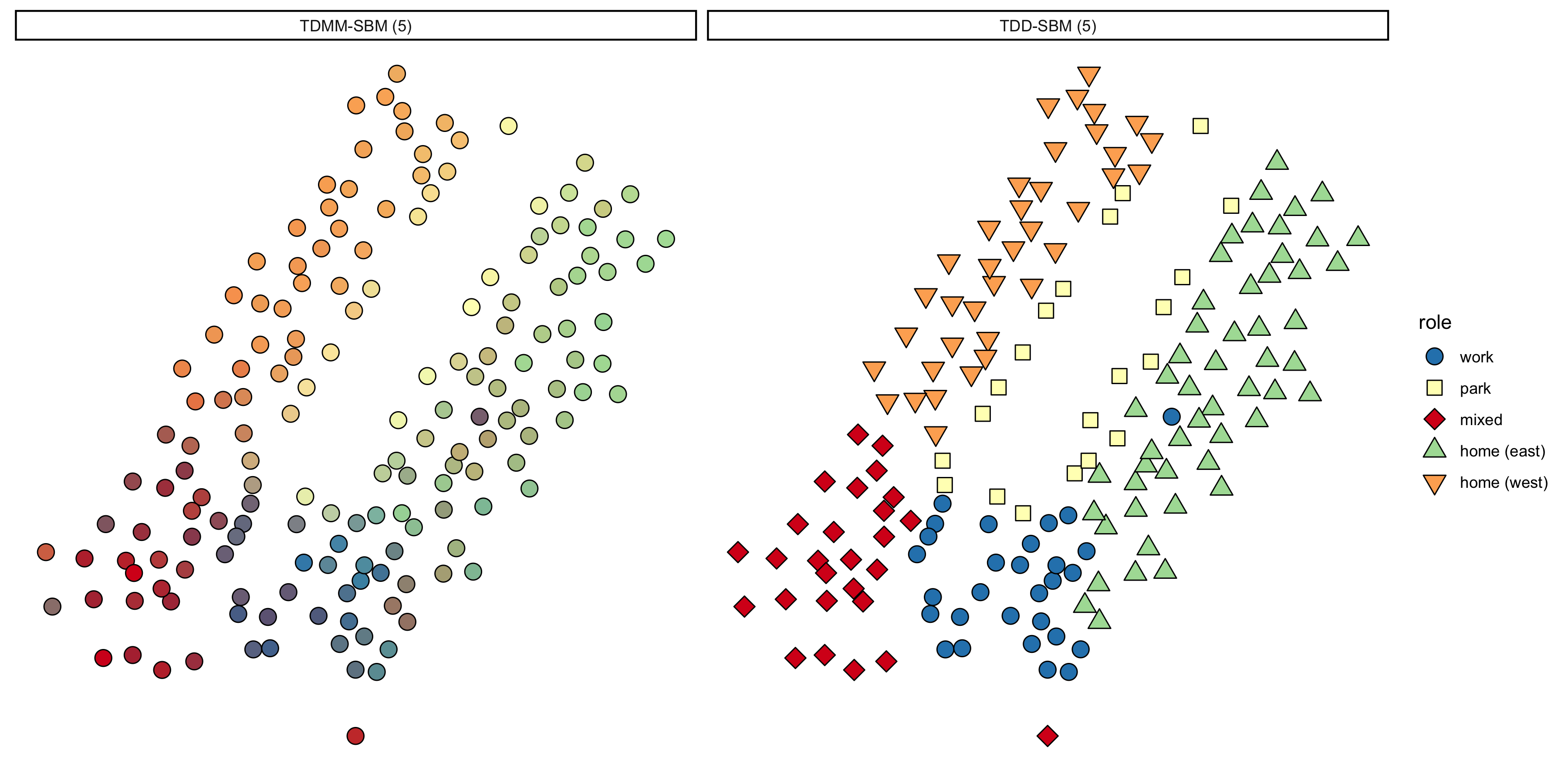}
\caption{Comparison of the blocks that we infer in the Manhattan subnetwork (i.e., the `Manhattan (home)' block in the right panel of Figure \ref{fig:NY_mixed_discrete}) of the New York City bicycle-sharing network using (left) a five-block TDMM-SBM and (right) a five-block TDD-SBM.} \label{fig:ny_plot_hm}
\end{figure}

We examine the New York City bicycle-sharing network on a scale that is smaller than the full system and fit models to the subset of stations and trips that lie within the Manhattan (home) block of the three-block TDD-SBM (see Figure \ref{fig:NY_mixed_discrete}) that we examine in Appendix \ref{appendix:nyc}. We refer to this subnetwork as the `Manhattan' network, even though it does not include all of the stations in the borough of Manhattan. This network consists of 166 stations and 256,840 trips. In Figure \ref{fig:ny_plot_hm}, we show our results for a five-block TDMM-SBM and five-block TDD-SBM on the Manhattan network. The area without stations in the middle of each panel of Figure \ref{fig:ny_plot_hm} is Central Park, which has stations on its perimeter but not in its interior.

The blocks that we infer using the five-block TDMM-SBM and five-block TDD-SBM outline similar subregions. These models yield block-membership parameters that capture the residential and commercial sections of the area much better than the three-block TDD-SBM and TDMM-SBM of the full New York City network. One can see this by comparing the five-block network results with the underlying zoning map for the area in Figure \ref{fig:ny_discrete_zones}. The stations in residential zones tend to have larger block-membership parameters for `home' blocks than for `work' blocks, and the opposite is true for stations in commercial zones. We label the five detected blocks as (clockwise from top left) `home (west)', `park', `home (east)', `work', and `mixed'. We base these labels on the land usage of the underlying areas and the time-dependent block-connectivity parameters $\hat{\omega}_{ght}$ (see Figure \ref{fig:ny_hm_omega}). The TDMM-SBM block assignments demarcate similar geographical subregions for the blocks as the TDD-SBM block assignments. However, when we use the TDMM-SBM, the subregions tend to blend together, rather than having sharp boundaries. (See Figure \ref{fig:ny_plot_hm}.)

We highlight the appearance of the `park' block, which we have not observed in previous examples and has distinctive behavior. The park block is similar to a residential block in terms of its spike in morning traffic to the work block and its spike in evening traffic from the work block, but it has distinct intra-block activity that peaks in the afternoon. The intra-block activity resembles weekend activity in the New York City bicycle-sharing system as a whole (see Figure \ref{fig:byhour}); this reflects leisure use of the bicycles. Bicycles near Central Park (which also places them near several major museums) are likely to be used by tourists and other non-commuters during the day for leisure or for travel to nearby attractions. 

In Figure \ref{fig:ny_hm_omega}, we show the values of the block-connectivity parameters $\hat{\omega}_{ght}$ for the five-block TDD-SBM and the five-block TDMM-SBM. Our estimates of $\hat{\omega}_{ght}$ for these models illustrate important differences in the behavior of different blocks that we can observe only by using a time-dependent model.\footnote{We obtain similar block identifications for this network using a discrete, directed, degree-corrected, time-independent SBM as we do from a TDD-SBM with the same number of blocks. We do not show the time-independent SBM results, but one can produce them using the code that we provide at \url{https://github.com/jcarlen/tdsbm_supplementary_material}.} We observe some overlap in the time-dependent behavior of blocks, so there is potential overfitting. For example, the home (east), home (west), and mixed blocks have traffic that is similar to that in blocks other than their own. However, examining the Manhattan network with models with fewer than five blocks do not cleanly distinguish the `park' block of stations from other residential stations.

\begin{figure}[H] \center 
\includegraphics[scale=.10]{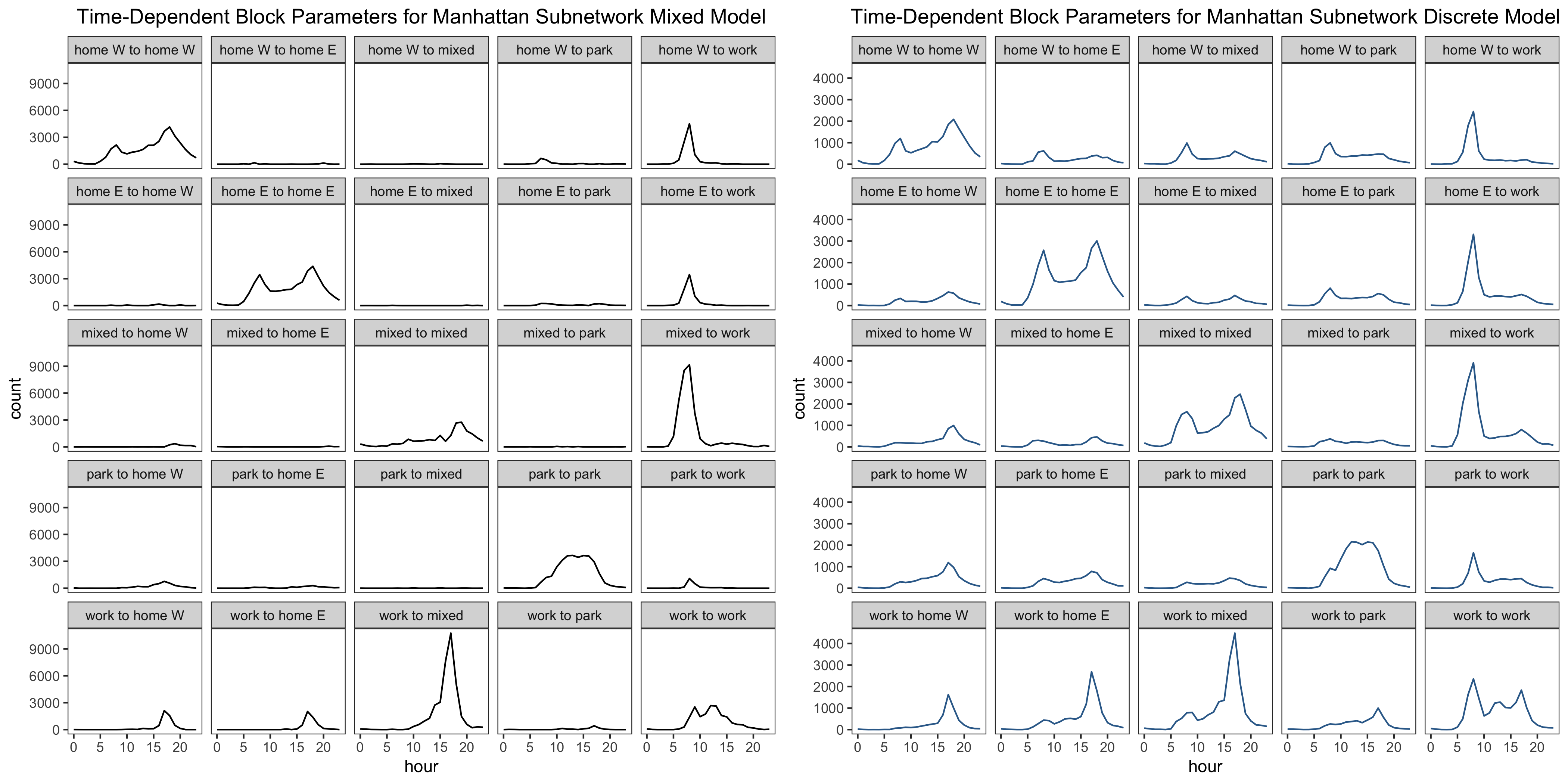}
\caption{Estimates of the time-dependent block-connectivity parameters $\hat{\omega}_{ght}$ in of the Manhattan subnetwork of the New York City network using (left) a TDMM-SBM with five blocks and (right) a TDD-SBM with five blocks.}
\label{fig:ny_hm_omega}
\end{figure}

One reason that our time-dependent SBMs for the Manhattan network perform better (with respect to detecting functionally meaningful blocks) than models that we apply to the entire New York City network (see Appendix \ref{appendix:nyc}) is the dependence of station-to-station trip counts on the distance between stations. Although our SBMs correct for the overall activity of each station, they do not normalize expected edge values by the distance between stations. In a small geographical area, such as the coverage areas of the Los Angeles and San Francisco networks, this is a reasonable choice, as all stations are within `biking distance' of each other. However, when examining a system as large as New York City's, the lack of distance correction weakens the functional groupings that we obtain with our time-dependent SBMs. Intra-block trips dwarf inter-block trips (see Figure \ref{fig:ny_omega}), and it seems more reasonable to interpret each block as its own ecosystem.


\section{Conclusions and Discussion}\label{Discussion}

We presented two types of time-dependent stochastic block models and used them to reveal interesting structures in urban bicycle-sharing networks. We formulated these SBMs to account both for degree heterogeneity and for a balance between cumulative in-degrees and out-degrees of bicycle stations over the course of a day. (The latter feature reflects a classical axiom of Ravenstein \cite{ravenstein1885} that every current of human mobility has an associated countercurrent \cite{barbosa2018human}.) Our SBMs group stations based on their activity over time, allowing us to identify them with home and work roles. Work stations are characterized by inflow from home stations in the morning and outflow to home stations in the afternoon and evening, and residential stations have the opposite characterization. It is also sometimes possible to identify other roles, such as a Manhattan park block that combines residential and leisure/touristic behavior.

We illustrated through case studies in Los Angeles, San Francisco, and New York City how our mixed-membership and discrete SBMs can provide complementary insights about transportation patterns in bicycle-sharing systems. We found that many bicycle-sharing stations in our three focal cities serve a mixture of roles, which we captured with our mixed-membership SBM. However, in some cases, we observed that discrete-membership SBMs that use fewer parameters provide a clearer picture of the usage patterns that are associated with each block. We also demonstrated the importance of time-dependence and degree correction in our models by comparing our results to time-independent and SBMs without degree correction.

We evaluated our block labels by comparing them to city zoning maps. The home--work structure that we detected generally aligns well with the underlying zones, but with important deviations near major transit hubs. It is common to evaluate the results of community detection by comparison with so-called `ground-truth' communities \cite{barbillon2017stochastic,fortunato2016community} (although it is crucial to encourage caution with respect to such evaluations \cite{peel2017ground}). The time-dependent commute flows that we detected with our SBMs enabled us to identify and label the functional roles of blocks of bicycle-sharing stations without a corresponding zoning map. In the future, it will be worthwhile to compare the traffic patterns that we detected to activity patterns from other mobility data, such as taxis, subway systems, e-scooters, and geo-tagged mobile apps~\cite{zhu18multi}.

There are several worthwhile ways to improve our models and algorithms. One valuable extension is to allow distributions other than the Poisson distribution for the numbers of trips between pairs of stations. Additionally, although it is a convenient simplification to assume independence between the number of trips that start each hour, it may be beneficial to relax this assumption. For example, they are not independent if there are stations that run out of bicycles at some point. As we have discussed at length, mixed-membership and discrete-membership SBMs can reveal different insights, as can examining SBMs with different numbers of blocks. This helps illustrate why it is important to consider model selection in greater depth. It is also worth considering other estimation techniques for both our mixed-membership and discrete-membership models, especially if one alters the assumption of conditionally independent Poisson edge distributions that we used to simplify our parameter estimation. Variational inference has been employed for many related models \cite{yang2011dsbm, matias2017dynsbm, matias2017semi, xing2010dmsbm}, and it is likely that it can also be used in our setting to improve performance (especially for large networks).

Another important direction for future work is the exploration of different methods for preprocessing data to include only the most important edges. The two most apparent ways to do this are (1) eliminating potentially insignificant edges by thresholding and (2) choosing time layers that reduce variance. The preferential-attachment model of Zhu et al.~\cite{zhu2013oriented} gives one possible approach for eliminating insignificant edges. The way that one splits the times of a day can improve both accuracy and computational efficiency by reducing the total number of parameters. For example, in the cities that we studied, bicycle trips occurred sporadically between 1 am and 5 am, so it may be desirable lump all of these time layers into one time interval to decrease the number of parameters by three and thereby decrease variance. There exist methods that aim to find suitable ways to segment time periods \cite{caceres2013temporal}, and trying to find the best ones to use in different situations is an active area of research.

Broadening our models to incorporate spatial data \cite{barthelemy2018morphogenesis} is another natural direction to build on our work. For example, it may enable one to identify functional blocks in geographically diffuse networks such the New York City bicycle-sharing network. The radiation, intervening-opportunities, and gravity models have had some success at modeling human mobility over various distances \cite{barbosa2018human}. These models put more weight on longer trips, and some mobility models take into account opportunities that lie between origin and destination locations. Some mobility models also possess statistical justification based on entropy arguments, and it is worthwhile to investigate methods to incorporate them into SBMs. Mobility models have already been incorporated into null models in time-dependent modularity objective functions in the work of Sarzynska et al. \cite{sarzynska2016null}, who found that radiation and gravity null models perform better than the usual Newman--Girvan null model (which is a variant of a configuration model \cite{fosdick17null}) for spatial networks. One can use similar ideas to incorporate spatial information into SBMs. The value of using spatial null models for bicycle-sharing systems has been examined previously \cite{austwick2013structure, aqil14tp19}, so this is a very interesting direction to pursue.

It is also worthwhile to use our time time-dependent SBMs and extensions of them to study other types of mobility data. One example is dockless vehicle-sharing networks, such as e-scooter-sharing programs. If we view the usage of stations as a proxy for a spatial function of demand for bicycles, then data from dockless systems may better approximate such a spatially varying function. One possibility is to partition a city into a grid (including comparing computations that use different levels of granularity) and measure the usage in each region, taking care to recognize irregularities from transit hubs. Depending on how heavily these systems are used in commuting, we may discover functional blocks other than ones for home `home' and `work'.  Looking even further forward, it will also be possible to tailor our methods to analyze multimodal transportation systems and other urban flows, which are particularly suitable for analysis using multilayer networks \cite{kivela2014,nutshell2019, natera2019multiplexbike,gallotti2019floworg}.


\section*{Acknowledgements}

We thank Brian Karrer and Mark Newman for allowing us to use and share their code for degree-corrected SBMs from \cite{karrer2011stochastic}. We thank Susan Handy's lab at UC Davis for useful discussions on contextualizing our work for transportation researchers and planners, David Kempe for introducing us to the paper \cite{caceres2013temporal}, and Michelle Feng and others in the networks journal club at UCLA for helpful comments. We also thank Phil Chodrow for technical assistance with setting up computations. CM and SSC thank NSF (DMS-1351860) for funding, and SSC also thanks NIGMS (R01 GM126556) and an NIH Ruth L. Kirschstein National Research Service Award (T32-GM008185) for funding. SW thanks NSF (CCF-1422795), ONR (N000141110719, N000141210834), DOD (W81XWH-15-1-0147), Intel STC-Visual Computing Grant (20112360), and Adobe Inc. for funding.


\break



\renewcommand\thefigure{A.\arabic{figure}}    
\renewcommand\thetable{A.\arabic{table}}    
\setcounter{figure}{0} 
\setcounter{table}{0}

\section*{Appendix}\label{appendix}


\subsection{Singular vectors for bicycle-sharing data}\label{appendix:svd}

{To inform our modeling approach}, we explore the imbalance between morning and evening activity by {station for each} network. We calculate the singular-value decomposition (SVD) of the matrices of in-degree and out-degree for each station by hour. To be explicit, entry $i,j$ of the matrix of in-degrees is equal to the total number of trips that arrive at station $i$ in hour $j$, and we construct the matrix of out-degrees analogously for departing trips. We show results for Los Angeles, San Francisco, and New York City in figures \ref{fig:la_svd}, \ref{fig:sf_svd}, and \ref{fig:ny_svd}, respectively. The first two principle components either strengthen both observed peaks or weaken one peak while strengthening the other. {Across the three data sets, the first two singular vectors explain between $88 \%$ and $97 \%$ percent} of the variation in the corresponding matrix, supporting the importance of peak morning and peak evening commutes for classifying stations.

\begin{figure}
[H]
\includegraphics[scale=.14]{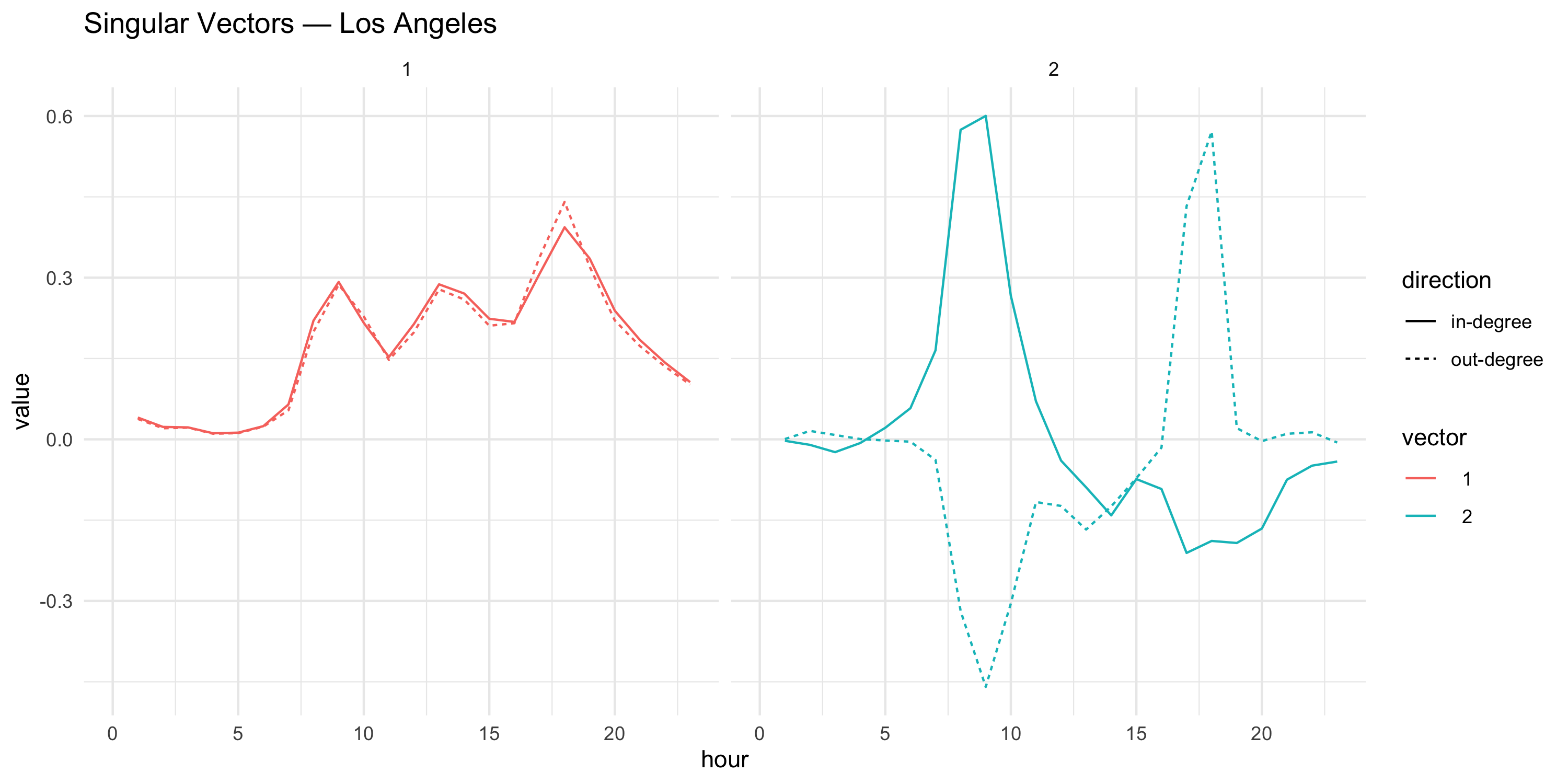}
\caption{The first two singular vectors of the in-degree matrix and the out-degree matrix of the downtown Los Angeles bicycle-sharing network.}\label{fig:la_svd}
\end{figure}

\begin{figure}[H]
\includegraphics[scale=.14]{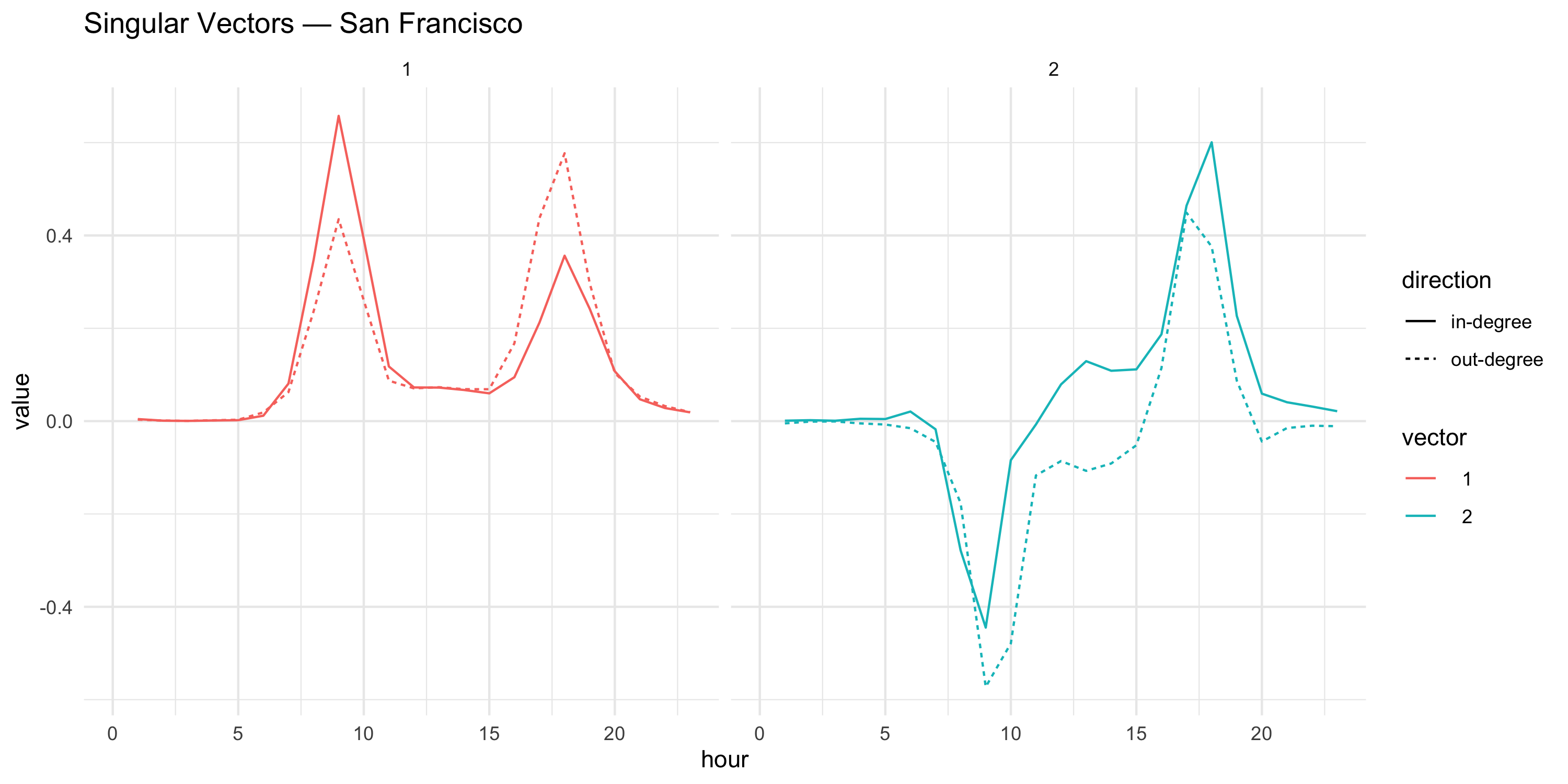}
\caption{The first two singular vectors of the in-degree matrix and the out-degree matrix of the San Francisco bicycle-sharing network.}\label{fig:sf_svd}
\end{figure}

\begin{figure}[H]
\includegraphics[scale=.14]{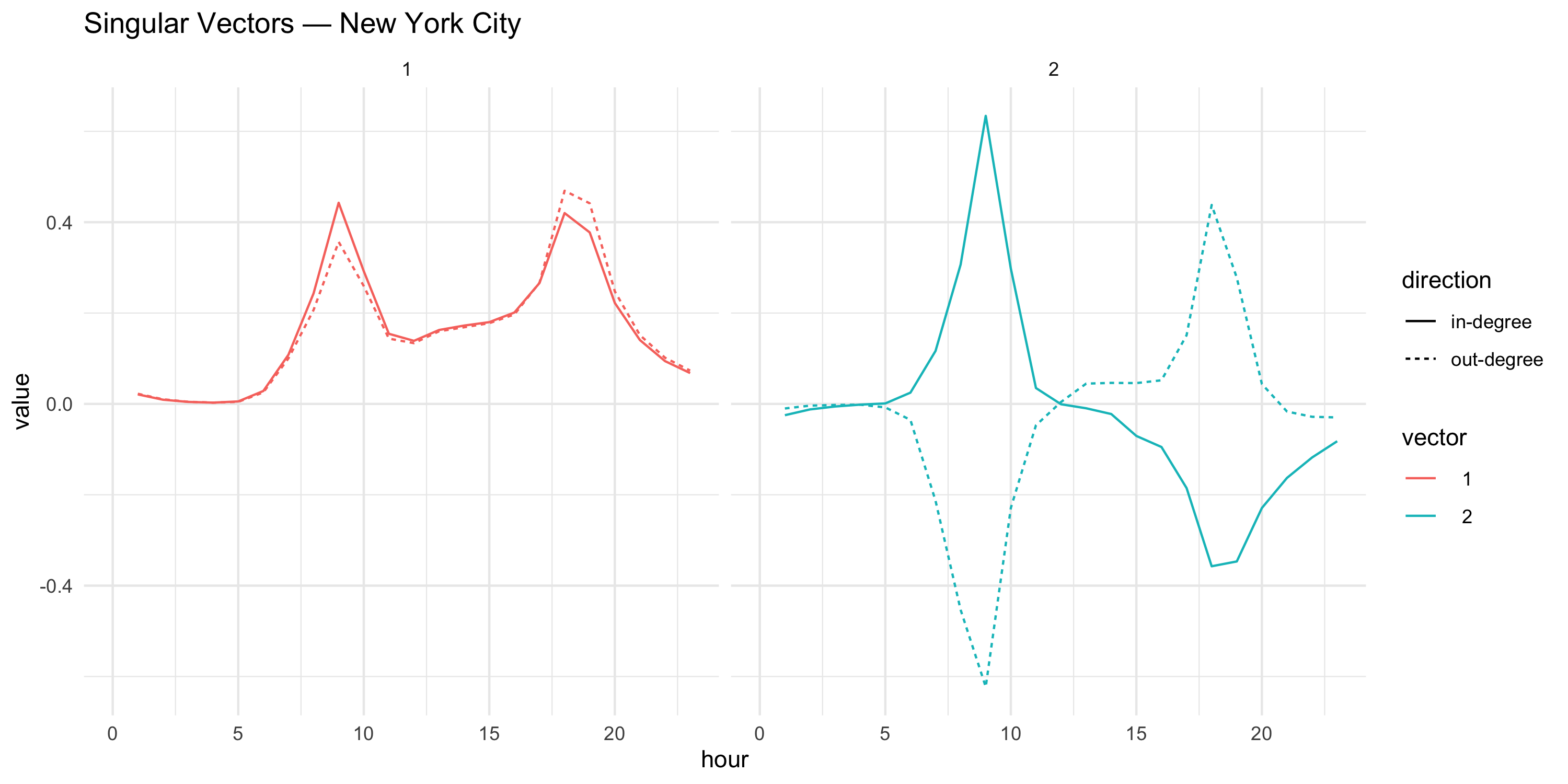}
\caption{The first two singular vectors of the in-degree matrix and the out-degree matrix of the New York City bicycle-sharing network.}\label{fig:ny_svd}
\end{figure}


\subsection{Proof that the expected node degrees of the MLE network of the TDD-SBM equal the observed node degrees}\label{appendix:node_deg} 

Suppose that we generate a network from the MLE of our time-dependent discrete-membership stochastic block model (TDD-SBM). We prove that the expected value of the degree of a node (i.e., the sum of its in-degree and out-degree over all time periods)  is the same as the degree of the node in the observed data. That is, we prove that $\sum_{j}\sum_{t=0}^{23} \left(\mu_{ijt} + \mu_{jit}\right) = k_i=\sum_j \sum_{t=0}^{23} A_{ijt}$.

Let $i$ and $j$ be nodes in the network, and let $t \in \{0,1,\ldots,23\}$ index the time layers. Additionally, recall that $g$ denotes block $g$ in the SBM and that $\kappa_g$ is the sum of the in-degrees and the out-degrees of all nodes in block $g$ over all time periods. We calculate 
\begin{align*}
{\sum_{j}\sum_{t=0}^{23} \left(\mu_{ijt} + \mu_{jit}\right)} &=\sum_{j}\sum_{t=0}^{23}\theta_i\theta_j(\omega_{g_ig_jt} + \omega_{g_jg_it})\\
	&=
  \frac{k_i}{\kappa_{g_i}}\sum_t \sum_h \sum_{j\in h}\frac{k_j}{\kappa_{h}}\left(m_{g_iht} + m_{hg_it}\right)\\
	&=
  \frac{k_i}{\kappa_{g_i}}\sum_{t}\sum_{h}\left(m_{g_iht} + m_{hg_it}\right)\\
	&=
  k_i\,.
\end{align*}

We are not aware of a relationship between the expected degrees and degrees of the observed data for our mixed-membership SBM. 


\subsection{{TDMM-SBM and TDD-SBM for the }New York City Bicycle-Sharing Network}\label{appendix:nyc}

In Figure \ref{fig:NY_mixed_discrete}, we compare our results from a three-block TDMM-SBM and a three-block TDD-SBM for the New York City bicycle-sharing network. In initial calculations, we found that a two-block TDD-SBM divides the network along the East River into a Manhattan block and Brooklyn block and that a two-block TDMM-SBM divides the network slightly farther north in Lower Manhattan. We do not show the outputs of the two-block models, but we {provide} the reproduction code, estimated parameters, and associated visualizations at \url{https://github.com/jcarlen/tdsbm_supplementary_material}.

\begin{figure}[H]
\includegraphics[scale=.14]{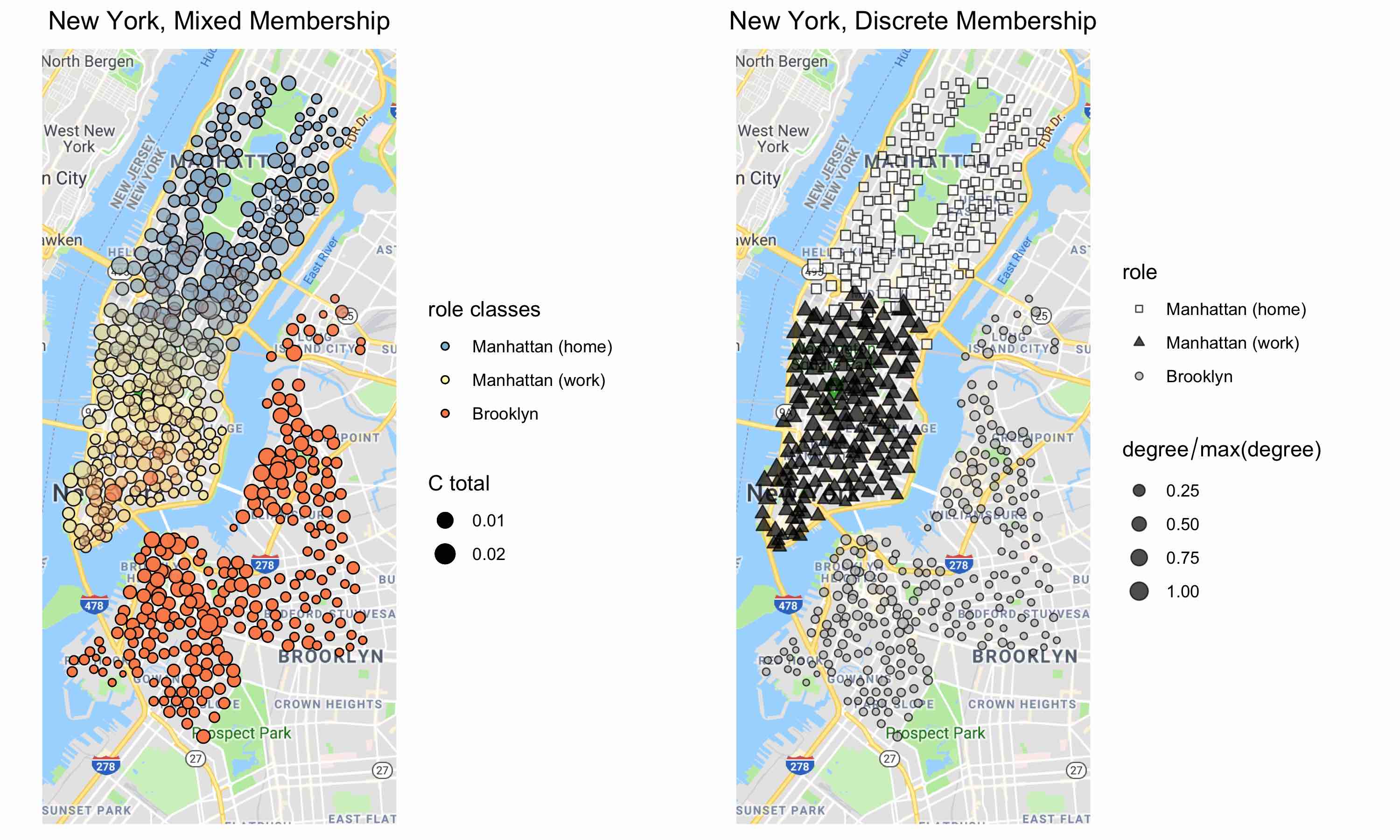}
\caption{Classification of New York City bicycle-sharing stations using (left) a three-block TDMM-SBM and (right) a three-block TDD-SBM. The sizes of the nodes take continuous values. In the left panel, we scale the area of each node based on the value of $\sum_gC_{ig}$. In the right panel, we scale the area of each node based on the sum of its in-degree and out-degree (divided by the maximum value of that sum).}\label{fig:NY_mixed_discrete}
\end{figure}

In Figure \ref{fig:ny_omega}, we compare the inferred inter-block traffic, as captured by the values of $\hat{\omega}_{ght}$, for the three-block TDMM-SBM and three-block TDD-SBM. We observe prominently that all intra-block traffic has two peaks and much higher hourly trip counts than inter-block traffic. The double peaks are reminiscent of the overall system activity in Figure \ref{fig:byhour}. This may be due in part to last-mile commuting, as we saw in the San Francisco bicycle-sharing network. However, for a system this geographically large, the double peaks and minimal inter-block traffic suggests that it is useful (and important) to consider each block as its own ecosystem. We also find strong similarity between results from our TDD-SBM and a three-block time-independent SBM for time-aggregated data for New York City (not shown), providing further evidence that our time-dependent SBMs are not capturing time-dependent roles for New York City when we consider the its entire bicycle-sharing system. Consequently, we choose the labels of these blocks (see Figures \ref{fig:NY_mixed_discrete} and \ref{fig:ny_omega}) based on the primary borough and zone type of each block's stations, as indicated in the zoning map for this part of New York City \cite{ny_zoning} in Figure \ref{fig:ny_discrete_zones}.

\begin{figure}[H]
\includegraphics[scale=.10]{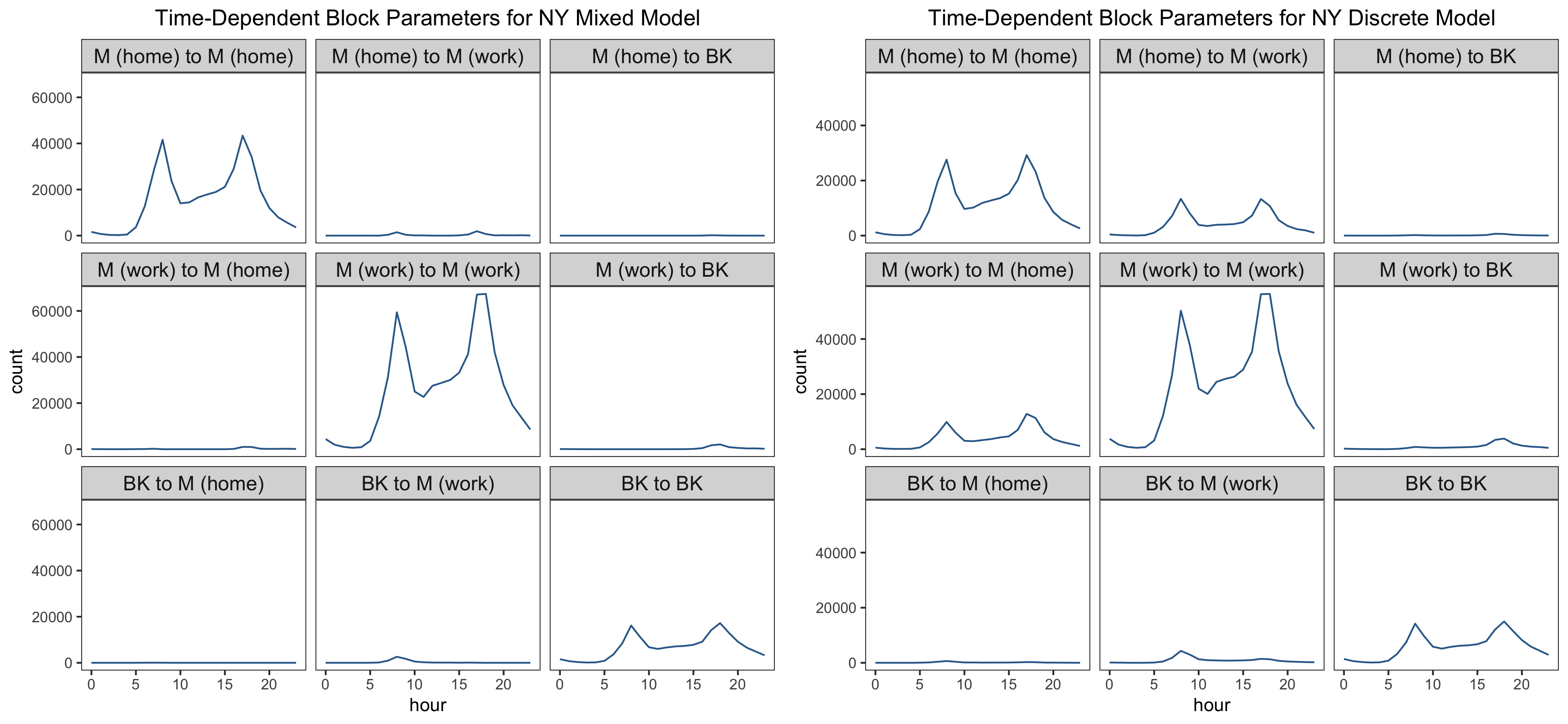}
\caption{Our estimates of the time-dependent block-connectivity parameters $\hat{\omega}_{ght}$ for the three-block TDMM-SBM and the three-block TDD-SBM for New York City. We use `M' to signify Manhattan and `BK' to signify Brooklyn.} \label{fig:ny_omega}
\end{figure}

\begin{figure}
\center 
\includegraphics[scale=.14]{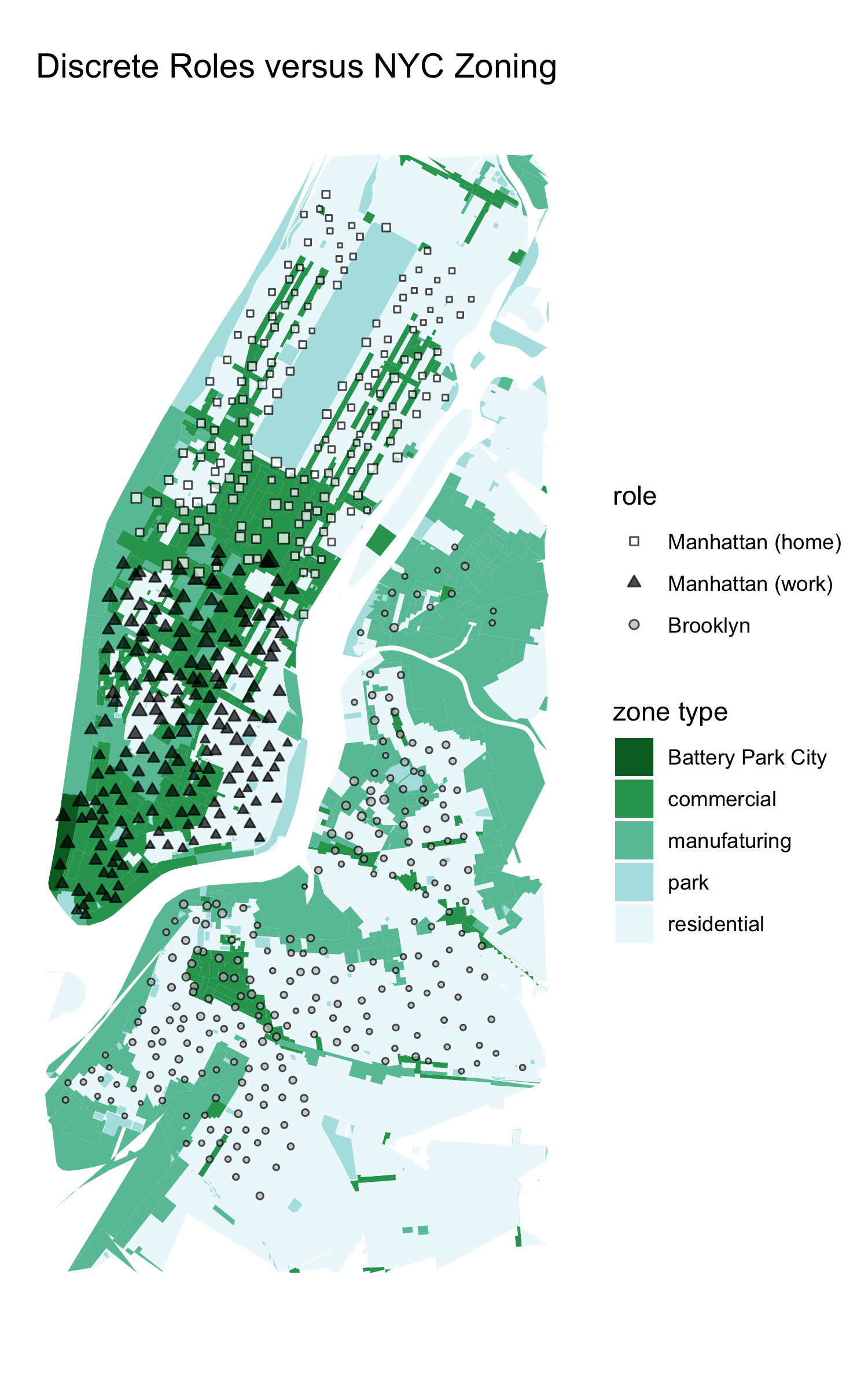}
\caption{The TDD-SBM station roles versus the coverage-area zoning map of New York City.}\label{fig:ny_discrete_zones}
\end{figure}

In Figure \ref{fig:ny_discrete_zones}, we illustrate that there is general overlap, although it is far from perfect, between (1) the Upper Manhattan (`home') block and residential areas or parks and between (2) the Lower Manhattan (`work') block and commercial or manufacturing areas. All stations in Brooklyn are in the third block, which contains mostly residential areas. We show only the blocks that we infer using our TDD-SBM in Figure \ref{fig:ny_discrete_zones}, but the same reasoning motivates our labels for the three-block TDMM-SBM. No block has exclusively commercial or residential areas.

We examined several time-dependent SBMs for New York City with more than three blocks to try to discover functional blocks, but we found that the blocks were still primarily geographical. In some cases, the TDMM-SBM with four or more blocks found functional divisions within smaller geographical areas (subdividing the blocks in Figure \ref{fig:NY_mixed_discrete}), but {none of our models} detected system-wide `home' or `work' blocks. See our code and figures at \url{https://github.com/jcarlen/tdsbm_supplementary_material} to fit and visualize time-dependent SBMs of the New York City network with a number of blocks other than three. 

Across our models of this network, the distance between pairs of stations is the feature that is associated most strongly with whether they belong to the same block. As we noted above, the detected blocks represent primarily geographical divisions, with most of the traffic occurring within blocks. Our time-dependent models do not detect functional blocks in New York City, in contrast to the results of applying them to the Los Angeles and San Francisco networks. We believe that this occurs because the New York City bicycle-sharing network is geographically larger than the San Francisco and Los Angeles bicycle-sharing networks. Our models only detect functional blocks if such divisions correlate with traffic between stations.
 

\subsection{Model Selection}\label{ModelSelection}

Although statistically rigorous model selection is outside the scope of our paper, we briefly compare the number of parameters in our mixed-membership SBM (TDMM-SBM) and discrete-membership SBM (TDD-SBM). This is valuable for considering model-selection criteria, such as the Akaike information criterion (AIC) and Bayesian information criterion (BIC), that penalize a model based on its number of parameters. For a network with $N$ nodes, $K$ blocks, and $T$ time layers, the number of parameters in the TDMM-SBM is 
\begin{align}\label{first}
    KN - K +  TK^2
\end{align} 
and the number of parameters in the TDD-SBM is
\begin{align}\label{second}
    2N - K + TK^2\,.
\end{align} 

The first term of \eqref{first} comes from the fact that each node in the TDMM-SBM has $K$ parameters ($C_{ig}$, with $g \in \{1,\ldots,K\}$) that express the strength of membership in each block. By contrast, each node in the TDD-SBM has one parameter for block membership and one degree-correction parameter. Therefore, given a value of $N$, the first term in \eqref{first} increases linearly with the number of blocks, whereas the corresponding term in \eqref{second} is fixed. Otherwise, formulas \eqref{first} and \eqref{second} are equivalent. The $-K$ term in each formula arises from the identifiability constraints in each model. As we described in Section \ref{Model}, these constraints are $\sum_i C_{ig}=1$ (for each block $g$) for the TDMM-SBM and $\sum_{i\in g}\theta_i = 1$ (for each block $g$) for TDD-SBM. The last term (i.e., $TK^2$) in each formula is the total number of $\omega_{ght}$ terms in the model (see Section \ref{TDMM-SBM}).

To illustrate these model-selection criteria, we give the unnormalized log-likelihoods and numbers of parameters ($N_p$) in TDMM-SBM and TDD-SMB of the Manhattan subnetwork (which has $N = 166$ nodes) with $2$--$6$ blocks (see Table \ref{ny_hm_llik_table}). In this example, the TDMM-SBM outperforms the TDD-SBM with respect to log-likelihood when the two models have the same number of parameters. This result makes sense because of the additional constraint of the TDD-SBM that each station must belong to exactly one block.

\begin{table}[ht!]
\centering
\begin{tabular}{rrrrrr}
& \textbf{TDMM-SBM} & & \textbf{TDD-SBM} & \\
  \hline
Blocks & $N_p$ & log-likelihood & $N_p$ & log-likelihood \\ 
  \hline
  
2 & 426 & $-260625$ & 426 & $-270809$ \\ 
3 & 711 & $-235162$ & 545 & $-254779$ \\ 
4 & 1044 & $-212295$ & 712 & $-236198$ \\ 
5 & 1425 & $-198489$ & 927 & $-222539$ \\ 
6 & 1854 & $-189670$ & 1190 & $-216468$ \\ 

   \hline
\end{tabular}
\caption{Comparison of log-likelihoods and numbers of parameters ($N_p$) in models of the Manhattan network, which has $N = 166$ nodes.}
\label{ny_hm_llik_table}
\end{table}

In Figure \ref{fig:AICvsblocks}, we plot the Akaike information criterion (AIC) \cite{akaike1974new}, which is given by 
\begin{align*}
    \text{AIC}=(2\times N_p) - (2\times \text{log-likelihood}) \,,
\end{align*} 
versus the number of blocks in the TDMM-SBM for the Los Angeles bicycle-sharing network with $2$--$10$ blocks. We consider models with a larger numbers of blocks in the Los Angeles network than we did in the Manhattan network because the former's smaller number of stations allows faster computations.

\begin{figure}[!ht] 
\center 
\includegraphics[scale=.28]{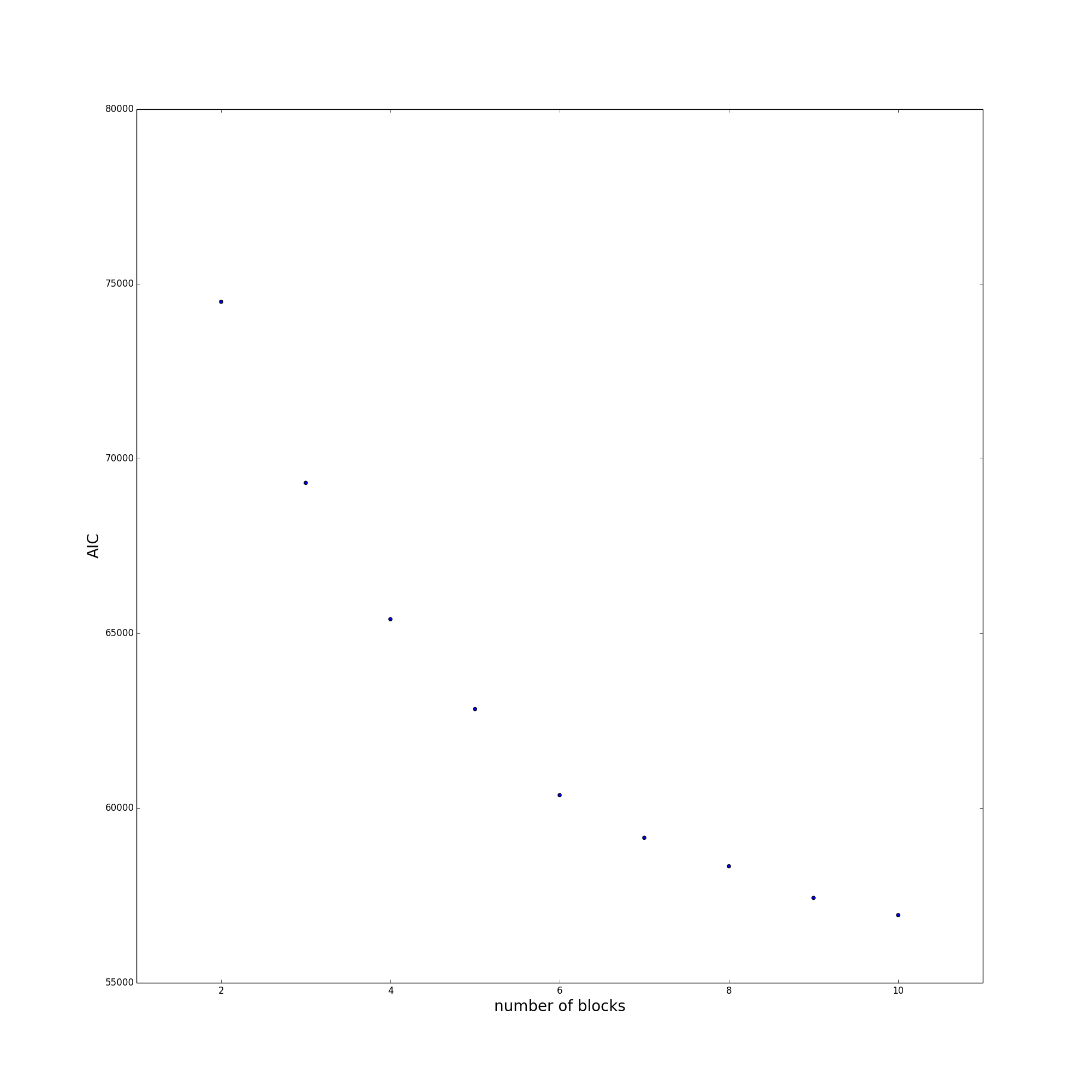}
\caption{Values of the Akaike information criterion for the MLE of the TDMM-SBM with $2$--$10$ blocks for the Los Angeles bicycle-sharing network.} 
\label{fig:AICvsblocks}
\end{figure}

The AIC is a cost function for comparing the relative qualities of statistical models. Given two models, one regards the model with a smaller AIC as the `better' model. The AIC takes into account both the likelihood and {the} complexity of a model, where one measures the latter based on the number of parameters in a model. In Figure \ref{fig:AICvsblocks}, we see that the AIC of the TDMM-SBM for the Los Angeles bicycle-sharing network decreases as we increase the number of blocks from $2$ to $10$. According to the AIC criterion, one would select the 10-block model. However, plots of the block-connectivity parameters $\hat{\omega}_{ght}$ for the TDMM-SBM of the Los Angeles network with seven or more blocks are noisier and no more informative than ones from the TDMM-SBM with fewer blocks. They suggest strongly that there is overfitting in those cases. This is in line with experiments by Funke and Becker {\cite{funke2019sbm}} who found that the AIC overestimates the number of blocks for several SBM variants.

Our time-dependent SBMs are incomplete descriptions of the data-generating process in the bicycle-sharing networks. For example, they do not account for the distances between stations. Accordingly, the AIC tends to recommend a model with more blocks to compensate for this incompleteness, but (as we just discussed) visual inspection reveals the likely overfitting that results from including too many blocks. Our goal is to gain insight into the bicycle-sharing systems, and we prefer models with fewer blocks that describe station roles in broad but interpretable term over less-interpretable models with more blocks but smaller AIC values. Choosing appropriate model-selection criteria for SBMs deserves additional consideration and is an active research area\cite{hu20cbic, wang17lbms, saldana17clbic, yan2016bayesian,yan2014model}. We leave such an investigation for future work.




\begin{thebibliography}{10}

\bibitem{barbosa2018human}
H.~Barbosa, M.~Barthelemy, G.~Ghoshal, C.~R. James, M.~Lenormand, T.~Louail,
  R.~Menezes, J.~J. Ramasco, F.~Simini, and M.~Tomasini, ``Human mobility:
  {M}odels and applications,'' {\em Physics Reports}, vol.~734, pp.~1--74,
  2018.

\bibitem{barthelemy2019}
M.~Barthelemy, ``The statistical physics of cities,'' {\em Nature Reviews
  Physics}, vol.~1, pp.~406--415, 2019.

\bibitem{schmidt2018ehp}
C.~Schmidt, ``Active travel for all? {T}he surge in public bike-sharing
  programs,'' {\em Environmental Health Perspectives}, vol.~128, Aug 2018.

\bibitem{nacto2017bikeshare}
``Bike share in the {U}.{S}.: 2017,'' tech. rep., {National Association of City
  Transportation Officials}, May 2018.
\newblock Available at \url{https://nacto.org/bike-share-statistics-2017/}.

\bibitem{etienne2014velib}
C.~Etienne and O.~Latifa, ``Model-based count series clustering for bike
  sharing system usage mining: {A} case study with the {V{\'e}lib'} system of
  {P}aris,'' {\em ACM Transactions on Intelligent Systems and Technology},
  vol.~5, no.~3, pp.~1--21, 2014.

\bibitem{austwick2013structure}
M.~Z. Austwick, O.~O'Brien, E.~Strano, and M.~Viana, ``The structure of spatial
  networks and communities in bicycle sharing systems,'' {\em PloS One},
  vol.~8, no.~9, p.~e74685, 2013.

\bibitem{munozmendez2018community}
F.~Munoz-Mendez, K.~Han, K.~Klemmer, and S.~Jarvis, ``Community structures,
  interactions and dynamics in {L}ondon's bicycle sharing network,'' in {\em
  Proceedings of the 2018 ACM International Joint Conference and 2018
  International Symposium on Pervasive and Ubiquitous Computing and Wearable
  Computers}, UbiComp '18, (New York, NY, USA), pp.~1015--1023, Association for
  Computing Machinery, 2018.

\bibitem{romanillos2016madrid}
G.~Romanillos and M.~Zaltz~Austwick, ``Madrid cycle track: {V}isualizing the
  cyclable city,'' {\em Journal of Maps}, vol.~12, no.~5, pp.~1218--1226, 2016.

\bibitem{romanillos2018pulse}
G.~Romanillos, B.~Moya-G{\'o}mez, M.~Zaltz-Austwick, and P.~J.
  Lam{\'\i}quiz-Daud{\'e}n, ``The pulse of the cycling city: {V}isualising
  {M}adrid bike share system {GPS} routes and cycling flow,'' {\em Journal of
  Maps}, vol.~14, no.~1, pp.~34--43, 2018.

\bibitem{wergin18gps}
J.~Wergin and R.~Buehler, ``Where do bikeshare bikes actually go?: {A}nalysis
  of capital bikeshare trips with {GPS} data,'' {\em Transportation Research
  Record}, vol.~2662, no.~1, pp.~12--21, 2017.

\bibitem{fishman2013bike}
E.~Fishman, S.~Washington, and N.~Haworth, ``Bike share: {A} synthesis of the
  literature,'' {\em Transport Reviews}, vol.~33, no.~2, pp.~148--165, 2013.

\bibitem{griffin2016lastmile}
G.~P. Griffin and I.~N. Sener, ``Planning for bike share connectivity to rail
  transit,'' {\em Journal of Public Transportation}, vol.~19, no.~2, pp.~1--22,
  2016.

\bibitem{ashqar2019counts}
H.~I. Ashqar, M.~Elhenawy, and H.~A. Rakha, ``Modeling bike counts in a
  bike-sharing system considering the effect of weather conditions,'' {\em Case
  Studies on Transport Policy}, vol.~7, no.~2, pp.~261--268, 2019.

\bibitem{natera2019multiplexbike}
L.~G. {Natera Orozco}, F.~Battiston, G.~I{\~n}iguez, and M.~Szell,
  ``Data-driven strategies for optimal bicycle network growth,'' {\em Royal
  Society Open Science}, vol.~7, p.~201130, 2020.

\bibitem{ashqar2020network}
H.~I. Ashqar, M.~Elhenawy, H.~A. Rakha, M.~Almannaa, and L.~House, ``Network
  and station-level bike-sharing system prediction: {A San Francisco Bay Area}
  case study,'' 2020.
\newblock arXiv:2009.09367.

\bibitem{shu2013redistribution}
J.~Shu, M.~Chou, Q.~Liu, C.~Teo, and I.-L. Wang, ``Models for effective
  deployment and redistribution of bicycles within public bicycle-sharing
  systems,'' {\em Operations Research}, vol.~61, no.~6, pp.~1346--1359, 2013.

\bibitem{prfrommer2014redistribution}
J.~Pfrommer, J.~Warrington, G.~Schildbach, and M.~Morari, ``Dynamic vehicle
  redistribution and online price incentives in shared mobility systems,'' {\em
  IEEE Transactions on Intelligent Transportation Systems}, vol.~15,
  pp.~1567--1578, Aug 2014.

\bibitem{singhvi2015redistribution}
D.~Singhvi, S.~Singhvi, P.~I. Frazier, S.~G. Henderson, E.~O. Mahony, D.~B.
  Shmoys, and D.~B. Woodard, ``Predicting bike usage for {New York City's} bike
  sharing system,'' in {\em Proceedings of the Association for the Advancement
  of Artificial Intelligence}, 2015.

\bibitem{forma2015redistribution}
I.~Forma, T.~Raviv, and M.~Tzur, ``A 3-step math heuristic for the static
  repositioning problem in bike-sharing systems,'' {\em Transportation Research
  Part B: Methodological}, vol.~71, pp.~230--247, 01 2015.

\bibitem{holme2012temporal}
P.~Holme and J.~Saram{\"a}ki, ``Temporal networks,'' {\em Physics Reports},
  vol.~519, no.~3, pp.~97--125, 2012.

\bibitem{holme2015}
P.~Holme, ``Modern temporal network theory: {A} colloquium,'' {\em The European
  Physical Journal B}, vol.~88, no.~9, 2015.

\bibitem{fortunato2016community}
S.~Fortunato and D.~Hric, ``Community detection in networks: {A} user guide,''
  {\em Physics Reports}, vol.~659, pp.~1--44, 2016.

\bibitem{kivela2014}
M.~Kivel{\"a}, A.~Arenas, M.~Barthelemy, J.~P. Gleeson, Y.~Moreno, and M.~A.
  Porter, ``Multilayer networks,'' {\em Journal of Complex Networks}, vol.~2,
  no.~3, pp.~203--271, 2014.

\bibitem{nutshell2019}
A.~Aleta and Y.~Moreno, ``Multilayer networks in a nutshell,'' {\em Annual
  Review of Condensed Matter Physics}, vol.~10, pp.~45--62, 2019.

\bibitem{Porter2018}
M.~A. Porter, ``{WHAT IS}... {A} multilayer network,'' {\em Notices of the
  American Mathematical Society}, vol.~65, no.~11, pp.~1419--1423, 2018.

\bibitem{lorrain1971struct}
F.~Lorrain and H.~White, ``Structural equivalence of individuals in social
  networks,'' {\em Journal of Mathematical Sociology}, vol.~1, pp.~49--80,
  1971.

\bibitem{rossi2015}
R.~A. Rossi and N.~K. Ahmed, ``Role discovery in networks,'' {\em IEEE
  Transactions on Knowledge and Data Engineering}, vol.~27, pp.~1112--1131,
  2015.

\bibitem{snijders1997sbm}
T.~A.~B. Snijders and K.~Nowicki, ``Estimation and prediction for stochastic
  block-structures for graphs with latent block structure,'' {\em Journal of
  Classification}, vol.~14, pp.~75--100, 1997.

\bibitem{nowicki2001sbm}
K.~Nowicki and T.~A.~B. Snijders, ``Estimation and prediction for stochastic
  blockstructures,'' {\em Journal of the American Statistical Association},
  vol.~96, no.~455, pp.~1077--1087, 2001.

\bibitem{karrer2011stochastic}
B.~Karrer and M.~E.~J. Newman, ``Stochastic blockmodels and community structure
  in networks,'' {\em Physical Review E}, vol.~83, no.~1, p.~016107, 2011.

\bibitem{peixoto2017bayesian}
T.~P. Peixoto, {\em Bayesian Stochastic Blockmodeling}, ch.~11, pp.~289--332.
\newblock John Wiley \& Sons, Ltd., 2019.

\bibitem{bazzi2016generative}
M.~Bazzi, L.~G.~S. Jeub, A.~Arenas, S.~D. Howison, and M.~A. Porter, ``A
  framework for the construction of generative models for mesoscale structure
  in multilayer networks,'' {\em Physical Review Research}, vol.~2, p.~023100,
  Apr 2020.

\bibitem{valdano2015analytical}
E.~Valdano, L.~Ferreri, C.~Poletto, and V.~Colizza, ``Analytical computation of
  the epidemic threshold on temporal networks,'' {\em Physical Review X},
  vol.~5, no.~2, p.~021005, 2015.

\bibitem{cranmer2015kantian}
S.~J. Cranmer, E.~J. Menninga, and P.~J. Mucha, ``Kantian fractionalization
  predicts the conflict propensity of the international system,'' {\em
  Proceedings of the National Academy of Sciences of the United States of
  America}, vol.~112, no.~38, pp.~11812--11816, 2015.

\bibitem{barbillon2017stochastic}
P.~Barbillon, S.~Donnet, E.~Lazega, and A.~Bar-Hen, ``Stochastic block models
  for multiplex networks: an application to a multilevel network of
  researchers,'' {\em Journal of the Royal Statistical Society: Series A
  (Statistics in Society)}, vol.~180, no.~1, pp.~295--314, 2017.

\bibitem{kobayashi2019nature}
T.~Kobayashi, T.~Takaguchi, and A.~Barrat, ``The structured backbone of
  temporal social ties,'' {\em Nature Communications}, vol.~10, no.~1, p.~220,
  2019.

\bibitem{papadopoulos2018network}
L.~Papadopoulos, M.~A. Porter, K.~E. Daniels, and D.~S. Bassett, ``Network
  analysis of particles and grains,'' {\em Journal of Complex Networks},
  vol.~6, no.~4, pp.~485--565, 2018.

\bibitem{mucha10multiplex}
P.~J. Mucha, T.~Richardson, K.~Macon, M.~A. Porter, and J.-P. Onnela,
  ``Community structure in time-dependent, multiscale, and multiplex
  networks,'' {\em Science}, vol.~328, no.~5980, pp.~876--878, 2010.

\bibitem{paul2016null}
S.~Paul and Y.~Chen, ``Null models and modularity based community detection in
  multi-layer networks,'' {\em arXiv:1608.00623}, 2016.

\bibitem{valles2016multilayer}
T.~Valles-Catala, F.~A. Massucci, R.~Guimera, and M.~Sales-Pardo, ``Multilayer
  stochastic block models reveal the multilayer structure of complex
  networks,'' {\em Physical Review X}, vol.~6, no.~1, p.~011036, 2016.

\bibitem{stanley2016clustering}
N.~Stanley, S.~Shai, D.~Taylor, and P.~J. Mucha, ``Clustering network layers
  with the strata multilayer stochastic block model,'' {\em IEEE Transactions
  on Network Science and Engineering}, vol.~3, no.~2, pp.~95--105, 2016.

\bibitem{yang2011dsbm}
T.~Yang, Y.~Chi, S.~Zhu, Y.~Gong, and R.~Jin, ``Detecting communities and their
  evolutions in dynamic social networks---{A} {B}ayesian approach,'' {\em
  Machine Learning}, vol.~82, no.~2, pp.~157--189, 2011.

\bibitem{zhang2017random}
X.~Zhang, C.~Moore, and M.~E.~J. Newman, ``Random graph models for dynamic
  networks,'' {\em The European Physical Journal B}, vol.~90, no.~10, p.~200,
  2017.

\bibitem{jeub2017}
L.~G.~S. Jeub, M.~W. Mahoney, P.~J. Mucha, and M.~A. Porter, ``A local
  perspective on community structure in multilayer networks,'' {\em Network
  Science}, vol.~5, no.~2, pp.~144--163, 2017.

\bibitem{xu2014dsbm}
K.~S. Xu and A.~O. Hero, ``Dynamic stochastic blockmodels for time-evolving
  social networks,'' {\em IEEE Journal of Selected Topics in Signal
  Processing}, vol.~8, no.~4, pp.~552--562, 2014.

\bibitem{matias2017dynsbm}
C.~Matias and V.~Miele, ``Statistical clustering of temporal networks through a
  dynamic stochastic block model,'' {\em Journal of the Royal Statistical
  Society: Series B (Statistical Methodology)}, vol.~79, no.~4, pp.~1119--1141,
  2017.

\bibitem{xing2010dmsbm}
E.~P. Xing, W.~Fu, and L.~Song, ``A state-space mixed membership blockmodel for
  dynamic network tomography,'' {\em The Annals of Applied Statistics}, vol.~4,
  no.~2, pp.~535--566, 2010.

\bibitem{ho2011dmsbm}
Q.~Ho, L.~Song, and E.~P. Xing, ``Evolving cluster mixed-membership blockmodel
  for time-varying networks,'' in {\em Proceedings of the 14th International
  Conference on Artificial Intelligence and Statistics} (G.~Gordon, D.~Dunson,
  and M.~Dudik, eds.), pp.~342--350, 2011.

\bibitem{rossetti2018dynamicsurvey}
G.~Rossetti and R.~Cazabet, ``Community discovery in dynamic networks: {A}
  survey,'' {\em ACM Computing Surveys}, vol.~51, no.~2, p.~35, 2018.

\bibitem{borgnat2011shared}
P.~Borgnat, P.~Abry, P.~Flandrin, C.~Robardet, J.-B. Rouquier, and E.~Fleury,
  ``Shared bicycles in a city: {A} signal processing and data analysis
  perspective,'' {\em Advances in Complex Systems}, vol.~14, no.~03,
  pp.~415--438, 2011.

\bibitem{rosvall2009infomap}
M.~Rosvall, D.~Axelsson, and C.~T. Bergstrom, ``The map equation,'' {\em The
  European Physical Journal Special Topics}, vol.~178, no.~1, pp.~13--23, 2009.

\bibitem{xie2018examining}
X.-F. Xie and Z.~J. Wang, ``Examining travel patterns and characteristics in a
  bikesharing network and implications for data-driven decision supports:
  {C}ase study in the {W}ashington {DC} area,'' {\em Journal of Transport
  Geography}, vol.~71, pp.~84--102, 2018.

\bibitem{akbarzadeh2018bike}
M.~Akbarzadeh, S.~S. Mohri, and E.~Yazdian, ``Designing bike networks using the
  concept of network clusters,'' {\em Applied Network Science}, vol.~3, no.~1,
  p.~12, 2018.

\bibitem{he2019divvytemporal}
M.~He, J.~Glasser, S.~Bhamidi, and N.~Kaza, ``Intertemporal community detection
  in bikeshare networks,'' 2019.
\newblock arXiv:1906.04582.

\bibitem{zhu18multi}
E.~Zhu, M.~Khan, P.~Kats, S.~Santosh~Bamne, and S.~Sobolevsky, ``Digital urban
  sensing: {A} multi-layered approach,'' 2018.
\newblock arXiv:1809.01280.

\bibitem{matias2017semi}
C.~Matias, T.~Rebafka, and F.~Villers, ``A semiparametric extension of the
  stochastic block model for longitudinal networks,'' {\em Biometrika},
  vol.~105, pp.~665--680, 06 2018.

\bibitem{citibikeData}
NYCBS, ``{Citi Bike System Data},'' 2017.
\newblock Available at
  \url{https://s3.amazonaws.com/tripdata/201610-citibike-tripdata.zip}; last
  checked 2019-05-05.

\bibitem{bayareaData}
{Bay Area Bike Share}, ``{Bay Area Bike Share Open Data},'' 2017.
\newblock Available at
  \url{https://s3.amazonaws.com/babs-open-data/babs_open_data_year_3.zip} last
  checked 2019-05-05; Archived main site:
  \url{https://web.archive.org/web/20170303021745/http://www.bayareabikeshare.com/open-data}.

\bibitem{metrobikeData}
{LA Metro}, ``{Metro Bike Share Trip Data},'' 2017.
\newblock Available at
  \url{https://bikeshare.metro.net/wp-content/uploads/2017/01/Metro_trips_Q4_2016.zip};
  last checked 2019-05-05; the version was last modified on 2018-09-17.

\bibitem{newman10networks}
M.~E.~J. Newman, {\em Networks}.
\newblock Oxford, UK: Oxford University Press, second~ed., 2018.

\bibitem{aqil14tp19}
A.~Taiyeb, ``{TP19}: {S}patial networks and human mobility: {A}n application of
  the intervening opportunities model to the {L}ondon cycle hire scheme.''
  Undergraduate Thesis, Department of Physics, University of Oxford (available
  at \url{ https://www.math.ucla.edu/~mason/research/aqil-TP19-final.pdf}),
  2014.

\bibitem{holland1983sbm}
P.~W. Holland, K.~B. Laskey, and S.~Leinhardt, ``Stochastic blockmodels:
  {F}irst steps,'' {\em Social Networks}, vol.~5, no.~2, pp.~109--137, 1983.

\bibitem{faust1992blockmodel}
K.~Faust and S.~Wasserman, ``Blockmodels: {I}nterpretation and evaluation,''
  {\em Social Networks}, vol.~14, no.~1--2, pp.~5--61, 1992.

\bibitem{airoldi2008mixed}
E.~M. Airoldi, D.~M. Blei, S.~E. Fienberg, and E.~P. Xing, ``Mixed membership
  stochastic blockmodels,'' {\em Journal of Machine Learning Research}, vol.~9,
  pp.~1981--2014, 2008.

\bibitem{latouche2011overlap}
P.~Latouche, E.~Birmel{\'e}, and C.~Ambroise, ``Overlapping stochastic block
  models with application to the {F}rench political blogosphere,'' {\em The
  Annals of Applied Statistics}, vol.~5, no.~1, pp.~309---336, 2011.

\bibitem{blockmodelsR}
J.-B. Leger, P.~Barbillon, and J.~Chiquet, {\em Blockmodels: Latent and
  stochastic block model estimation by a `V-EM' algorithm}, 2020.
\newblock R package version 1.1.4.

\bibitem{zhu2013oriented}
Y.~Zhu, X.~Yan, and C.~Moore, ``Oriented and degree-generated block models:
  {G}enerating and inferring communities with inhomogeneous degree
  distributions,'' {\em Journal of Complex Networks}, vol.~2, no.~1, pp.~1--18,
  2013.

\bibitem{stan2017jss}
B.~Carpenter, A.~Gelman, M.~Hoffman, D.~Lee, B.~Goodrich, M.~Betancourt,
  M.~Brubaker, J.~Guo, P.~Li, and A.~Riddell, ``Stan: {A} probabilistic
  programming language,'' {\em Journal of Statistical Software}, vol.~76,
  no.~1, pp.~1--32, 2017.

\bibitem{rstan2018}
{Stan Development Team}, ``{RStan}: {T}he {R} interface to {Stan},'' 2018.
\newblock R package version 2.18.2.

\bibitem{kernighanlin70}
B.~W. Kernighan and S.~Lin, ``An efficient heuristic procedure for partitioning
  graphs,'' {\em Bell System Technical Journal}, vol.~49, no.~2, pp.~291--307,
  1970.

\bibitem{r18}
{R Core Team}, {\em R: A Language and Environment for Statistical Computing}.
\newblock R Foundation for Statistical Computing, Vienna, Austria, 2018.

\bibitem{eddelbuettel11rcpp}
D.~Eddelbuettel and R.~Fran\c{c}ois, ``{Rcpp}: Seamless {R} and {C++}
  integration,'' {\em Journal of Statistical Software}, vol.~40, no.~8,
  pp.~1--18, 2011.

\bibitem{giorgi2018ppsbm}
D.~Giorgi, C.~Matias, T.~Rebafka, and F.~Villers, {\em ppsbm: Clustering in
  Longitudinal Networks}.
\newblock R package version 0.2.2. Available at
  \url{https://CRAN.R-project.org/package=ppsbm}.

\bibitem{vavrek2011fossil}
M.~J. Vavrek, ``fossil: palaeoecological and palaeogeographical analysis
  tools,'' {\em Palaeontologia Electronica}, vol.~14, no.~1, p.~1T, 2011.
\newblock R package version 0.4.0. Available at
  \url{https://palaeo-electronica.org/2011_1/238/index.html}.

\bibitem{la_zoning}
{Department of City Planning, City of Los Angeles}, ``{City of Los Angeles GIS
  Data: Zoning},'' 2015.
\newblock Available at \url{http://planning.lacity.org/}.

\bibitem{la_zoning_dict}
{Department of City Planning, City of Los Angeles}, ``Generalized summary of
  zoning regulations of city of {Los Angeles},'' January 2006.
\newblock Available at
  \url{https://planning.lacity.org/zone_code/Appendices/sum_of_zone.pdf}.

\bibitem{pubtranLAT2015}
D.~Weikel and S.~Karlamangla, ``San {F}rancisco residents relying less on
  private automobiles,'' {\em Los Angeles Times}, February 23 2015.

\bibitem{ravenstein1885}
E.~J. Ravenstein, ``The laws of migration,'' {\em Journal of the Statistical
  Society}, vol.~46, pp.~167--235, 1885.

\bibitem{peel2017ground}
L.~Peel, D.~B. Larremore, and A.~Clauset, ``The ground truth about metadata and
  community detection in networks,'' {\em Science Advances}, vol.~3, no.~5,
  p.~e1602548, 2017.

\bibitem{caceres2013temporal}
R.~S. Caceres and T.~Berger-Wolf, ``Temporal scale of dynamic networks,'' in
  {\em Temporal Networks} (P.~Holme and J.~Saram\"{a}ki, eds.), (Heidelberg,
  Germany), pp.~65--94, Springer-Verlag, 2013.

\bibitem{barthelemy2018morphogenesis}
M.~Barthelemy, {\em Morphogenesis of Spatial Networks}.
\newblock Cham, Switzerland: Springer International Publishing, 2018.

\bibitem{sarzynska2016null}
M.~Sarzynska, E.~A. Leicht, G.~Chowell, and M.~A. Porter, ``Null models for
  community detection in spatially embedded, temporal networks,'' {\em Journal
  of Complex Networks}, vol.~4, no.~3, pp.~363--406, 2016.

\bibitem{fosdick17null}
B.~K. Fosdick, D.~B. Larremore, J.~Nishimura, and J.~Ugander, ``Configuring
  random graph models with fixed degree sequences,'' {\em SIAM Review},
  vol.~60, no.~2, pp.~315--355, 2018.

\bibitem{gallotti2019floworg}
R.~Gallotti, G.~Bertagnolli, and M.~De~Domenico, ``Unraveling the hidden
  organisation of urban systems and their mobility flows,'' {\em The European
  Physical Journal --- Data Science}, vol.~10, p.~3, 2021.

\bibitem{ny_zoning}
{Department of City Planning, New York City}, ``{NYC} zoning districts,'' March
  2018.
\newblock Available at
  \url{https://www1.nyc.gov/site/planning/data-maps/open-data/dwn-gis-zoning.page}.

\bibitem{akaike1974new}
H.~Akaike, ``A new look at the statistical model identification,'' in {\em
  Selected Papers of Hirotugu Akaike}, pp.~215--222, Heidelberg, Germany:
  Springer-Verlag, 1974.

\bibitem{funke2019sbm}
T.~Funke and T.~Becker, ``Stochastic block models: A comparison of variants and
  inference methods,'' {\em PLoS ONE}, vol.~14, no.~4, p.~e0215296, 2019.

\bibitem{hu20cbic}
J.~Hu, H.~Qin, T.~Yan, and Y.~Zhao, ``Corrected bayesian information criterion
  for stochastic block models,'' {\em Journal of the American Statistical
  Association}, vol.~115, no.~532, pp.~1771--1783, 2020.

\bibitem{wang17lbms}
Y.~X.~R. Wang and P.~J. Bickel, ``Likelihood-based model selection for
  stochastic block models,'' {\em The Annals of Statistics}, vol.~45, no.~2,
  pp.~500--528, 2017.

\bibitem{saldana17clbic}
D.~F. Salda{\~n}a, Y.~Yu, and Y.~Feng, ``How many communities are there?,''
  {\em Journal of Computational and Graphical Statistics}, vol.~26, no.~1,
  pp.~171--181, 2017.

\bibitem{yan2016bayesian}
X.~Yan, ``Bayesian model selection of stochastic block models,'' in {\em
  Proceedings of the 2016 IEEE/ACM International Conference on Advances in
  Social Networks Analysis and Mining}, pp.~323--328, IEEE Press, 2016.

\bibitem{yan2014model}
X.~Yan, C.~Shalizi, J.~E. Jensen, F.~Krzakala, C.~Moore, L.~Zdeborov{\'a},
  P.~Zhang, and Y.~Zhu, ``Model selection for degree-corrected block models,''
  {\em Journal of Statistical Mechanics: Theory and Experiment}, vol.~2014,
  no.~5, 2014.

\end{thebibliography}



\end{document}